\documentclass[fleqn,usenatbib]{mnras}
\usepackage{hyperref}
\hypersetup{draft}

\usepackage[T1]{fontenc}
\usepackage{ae,aecompl}

\usepackage{graphicx}	
\usepackage{amsmath}	
\usepackage{amssymb}	

\newcommand{\dd}{ {\rm d} }

\title[Galactic chemical evolution including dust]{Chemical evolution of galaxies: emerging dust and the different gas phases in a new multiphase code}

\author[Mill{\'a}n-Irigoyen, Moll{\'a} \& Ascasibar.]{
I. Mill{\'a}n-Irigoyen$^{1}$,\thanks{E-mail: iker.millan@ciemat.es}
M. Moll{\'a},$^{1}$\thanks{E-mail: mercedes.molla@ciemat.es}
and Y. Ascasibar$^{2,3}$\thanks{E-mail: yago.ascasibar@uam.es}
\\
$^{1}$ Departamento de Investigaci{\'o}n B{\'a}sica, CIEMAT, Av. Complutense 40,E-28040, Madrid, Spain\\
$^{2}$ Departamento de F{\'i}sica Te{\'o}rica, Universidad Aut\'{o}noma de Madrid, E-28049, Cantoblanco (Madrid), Spain\\
$^{3}$ Astro-UAM, Unidad Asociada CSIC, Universidad Aut\'{o}noma de Madrid, E-28049, Cantoblanco (Madrid), Spain\\
}

\date{Accepted XXX. Received YYY; in original form ZZZ}

\pubyear{2019}

\begin{document}
\label{firstpage}
\pagerange{\pageref{firstpage}--\pageref{lastpage}}
\maketitle

\begin{abstract}
Dust plays an important role in the evolution of a galaxy, since it is one of the main ingredients for efficient star formation. Dust grains are also a sink/source of metals when they are created/destroyed, and, therefore, a self-consistent treatment is key in order to correctly model chemical evolution. In this work, we discuss the implementation of dust physics into our current multiphase model, which also follows the evolution of atomic, ionised and molecular gas.
Our goal is to model the conversion rates among the different phases of the interstellar medium, including the creation, growth and destruction of dust, based on physical principles rather than phenomenological recipes inasmuch as possible. We first present the updated set of differential equations and then discuss the results. We calibrate our model against observations of the Milky Way Galaxy  and compare its predictions with extant data. Our results are broadly consistent with the observed data for intermediate and high metallicities, but the models tend to produce more dust than observed in the low metallicity regime.
\end{abstract}




\section{Introduction}\label{sec:Intro}

Until the mid-$\mathrm{20}^{\mathrm{th}}$ century, cosmic dust was, for most astronomers, a foreground that complicated the observation of distant objects. When the first observations in the infrared became available, the chemical composition and distribution of dust in the interstellar medium (ISM) could be measured, and theories about its role in the evolution of galaxies emerged. Nowadays, it is widely known that interstellar dust is made up of very small solid particles (grains of $\la 1~\mu\mathrm{m}$ in size) of carbon, silicates and ices from other chemical species. It is also estimated that the entire mass constituting the ISM represents $\sim 10$\% of the total baryonic mass of a galaxy, of which $\sim 99$\% corresponds to gas, with only $\sim$1\% attributed to dust.

The so-called "dust cycle" consists  essentially on the following processes \citep[see][hereafter Z08]{Zhukovska2008}, c.f. \citep{Asano_2013, Accretion_Gioannini}:
\begin{enumerate}
    \item Ejection from stars and injection into the ISM.
    \item Evolution: production in the ISM, growth and destruction.
    \item Star formation: molecular cloud formation and shielding.
\end{enumerate}

Most theoretical models of dust formation are based on the condensation of metallic elements from their gaseous phase in a solid state through nucleation processes \citep{Nucleation_theory_historic,Todini_Ferrara,Nanni_2014}. For the condensation to happen, the partial pressure in the gas has to be greater than the vapour pressure in the condensed phase \citep{Valiante09}. The size of the grains basically depends upon three parameters related to the thermodynamic conditions: the gas temperature ($\rm T_{g}$), the gas pressure ($\rm P_{g}$) and the condensation temperature of the involved chemical species ($\rm T_{c}$). 

Nucleation  occurs in a range of $\rm P_{g}$ between $10^{-8}$ and $10^{5}$\,Pa, with $\rm T_{g} \le 1800$\,K, and these restrictive conditions limit the possible sources of dust grains observed in the ISM. There is a consensus that the major sources of dust in the Universe are stars of low and intermediate mass ($\sim 0.8\,\rm {M_{\sun}} \le m \le 8\,M_{\sun}$) having evolved through the asymptotic giant branch i.e., AGB stars \citep[Z08,][]{ AGB_Z_10_-3, AGB_Z_8_10_-3,   AGB_Z_3_10_-4, Nanni_2013, Dust_SMC, AGB_Z_1.8_10_-2, Ginolfi_2017}, as well as massive stars ($\rm m> 8\,M_{\sun}$) when they die as core-collapse type II SNe \citep[CC-SNe;][]{Todini_Ferrara,   Schneider2004, Nozawa2007, Bianchi_Schneider, Cherchneff2009,  Cherchneff2010, Marassi2014, SN_cc_dust, Sarangi2015, Bocchio2016,  Ginolfi_2017}.

It has been argued \citep[see e.g.][]{Morgan_Edmunds03, Valiante09,  Accretion_Gioannini} that other sources are necessary in order to reproduce the amounts of dust observed in the Universe, and many candidates have been proposed, such as SNe type Ia \citep{Nozawa2011}, red giant branch (RGB) stars \citep{Origlia2010}, luminous blue variable (LBV) stars (Z08), Wolf-Rayet (WR) stars \citep{Zhekov2014,Cherchneff2015} or even Population III stars  passing through their red supergiant (RSG) phase \citep{SN_cc_dust}.

After nucleation, dust undergoes different evolutionary processes in the ISM.  Some of these processes change the dust mass, such as growth by accretion \citep[e.g.][]{Inoue_2011, Zhukovska2014, Zhukovska2016,   Zhukovska2018} or its destruction by thermal or chemical sputtering, while others modify its properties (e.g. the shattering or fragmentation by grain collisions, which can affect the size distribution). Eventually, dust grains are incorporated into the next generations of stars (astration) or destroyed by either thermal pulverisation produced by the reverse shock in SN explosions \citep[e.g.][]{Bianchi_Schneider,  Nozawa2011} or the radiation field surrounding very luminous objects \citep[e.g.][]{Dust_destruction_Radiative_Torque}.

Despite the insignificance in the mass budget of a galaxy, the importance of the dust component can hardly be overstated.  Besides its impact on the observed spectral energy distribution (through the absorption, emission and scattering of photons at different wavelengths), the presence of dust grains plays a central role in the chemical evolution of any galactic region, as well as of the galaxy as a whole, since dust regulates the creation of molecules in the ISM and the shielding of molecular clouds from ultraviolet (UV) photons.

Over the years, several theoretical models have appeared that account for the abundance of one or more dust species simultaneously with the chemical elements in the gas phase of the ISM \citep[e.g.][]{Dwek_98, Calura_2008, Inoue_2011, Asano_2013, Zhukovska2014, Accretion_Gioannini, Zhukovska2016, Zhukovska2018}.

Building upon these studies, the main aim of the present work is to discuss the implementation of dust physics into a multiphase chemical evolution model.  This represents a significant improvement with respect to previous works; in some of the latter \citep[e.g.][]{AGB_gas_ejecta, Molla_2016, Molla_2017}, the availability of molecular gas is one of the main ingredients of the star formation process, but the dust phase is not included. In others, the dust is taken into account, but only crudely modelled, with the assumption that the dust mass scales approximately linearly with the metal content, according to some proportionality constant that was fitted as a phenomenological parameter.  Here, we will describe and test a physically motivated formalism that provides a much more realistic (and significantly more constrained) parametrization, and opens the possibility of testing the model results against dust-related observational data, increasing the predictive power of the models and helping to break the degeneracies intrinsic to chemical evolution studies.

Our new model is fully described in Section~\ref{Sec:Model}, where we explain in detail the equations that determine the transfer of mass between the different phases and the precise meaning of the small number of free parameters that appear only in the dust physical or phenomenological prescriptions.  After deriving our system of coupled differential equations, we can obtain the time evolution of the gas (including diffuse, ionised and molecular phases), stars, metals and dust densities in a given galaxy.  Including a careful treatment of the dust life cycle based on first principles is particularly important in relation to the phase balance and the star formation processes, since the main channel for the creation of hydrogen molecules is the condensation onto the surface of dust grains.

The first results of the model are presented in Section~\ref{sec:Results}, where we discuss the time evolution of a fiducial configuration that is representative of a Milky Way Galaxy (MWG) or solar neighbourhood-type model.  After verifying that we are able to successfully reproduce the local environment, we study the effect of varying the dust-related parameters (destruction timescale, cloud density, and sticking coefficient) and of changing the main inputs of the model (the total mass surface density and the infall timescale), which define a given size of galaxy, in order to disentangle their contribution to the evolution and the final outcome of the model. Finally, we compare the results of a model grid covering a broad range of parameter values with a set of observational data of dust abundances in external galaxies available in the literature.  Our man conclusions are summarised in Section~\ref{sec:Conclusions}.

\section{Model Description}\label{Sec:Model}

The main aim of chemical evolution models is to describe the temporal evolution of different chemical species within a galaxy,in order to confront the predicted behaviour with that observed. In our model, we consider that galaxies have a multiphase structure that includes neutral, ionised and molecular gas phases in the ISM, and we follow the abundance of hydrogen, helium, metals and dust. The equations that determine the evolution of each component will be discussed in subsections ~\ref{Subsec:Gas}, \ref{Subsec:Stars} and ~\ref{Subsec:Dust}. In order to compare our theoretical predictions with observations, the mass content of all phases is expressed in terms of surface density.

\begin{table}
\begin{center}
\caption{Glossary of symbols associated with each ISM phase}	\begin{tabular}{ll}
\hline
Symbol & Component \\
\hline
$\Sigma_{\mathrm{total}}$ & Total mass surface density, $ \Sigma_{\mathrm{total} } = \Sigma_* + \Sigma_{\mathrm{ISM}} $ \\
$\Sigma_*$ & Stars and stellar remnants \\
$\Sigma_{\mathrm{ISM} }$ & Interstellar medium, $\Sigma_{\mathrm{ISM} } = \Sigma_{\mathrm{gas} } + \Sigma_{\mathrm{dust} } $ \\
\hline
$\Sigma_{\mathrm{gas} }$ & Gas, $\Sigma_{\mathrm{gas}} = \Sigma_{\mathrm{i}} + \Sigma_{\mathrm{a}} + \Sigma_{\mathrm{m}} = \Sigma_X + \Sigma_Y + \Sigma_Z $ \\
$\Sigma_{\mathrm{i}}$ & Warm ionised medium \\
$\Sigma_{\mathrm{a}}$ & Diffuse atomic gas \\
$\Sigma_{\mathrm{m}}$ & Molecular clouds \\
$\Sigma_X$ & Hydrogen \\
$\Sigma_Y$ & Helium \\
$\Sigma_Z$ & Metals in the gas phase \\
\hline
$\Sigma_{\mathrm{d}}$ & Dust \\
\hline
			\end{tabular}
		\label{tab_components}
	\end{center}
\end{table}

A summary of the  phase-dependent notation adopted herein is provided in Table~\ref{tab_components}. The total mass of each galaxy is defined as the sum of the stellar and ISM mass.  This ISM, in turn,  is composed of warm ionised gas, diffuse atomic gas, and molecular clouds.  Within the gas, metals appears as a consequence of the evolution and death of stars. Thus, its composition is $\rm X+Y+Z$, where, as usual, $\rm X$ is the mass abundance of $\rm H$, $\rm Y$ is the mass abundance of $\rm He$ and $\rm Z$ is the mass abundance of all the remaining elements.  In our scenario, we start with a disk of zero total mass, $\rm \Sigma_{total}(t=0)=0$, and consider infall of gas from outside with primordial abundances \citep{coc12} that replenishes the disk galaxy. At the end of our runs \citep[13.2\,Gyr as in][]{AGB_gas_ejecta}, a disk remains with characteristics consistent with those observed in nearby spirals. 
\subsection{Gas}\label{Subsec:Gas}

Since the ISM in a galaxy is not homogeneous, it is indispensable to model it as a multiphase structure.  For that purpose, we have considered: a) an ionised phase with a temperature of $\sim 10^{4}$~K, b) an atomic component whose temperature is $\sim 100$~K, and c) a molecular phase with a temperature of $\sim 10$~K.  These phases interchange mass with each other during the evolution of the region under study.  The processes involved are: the photoionisation of atoms, the recombination of electrons with ions, the conversion of atomic hydrogen into molecular hydrogen and the destruction of the latter due to the photodissociation caused by UV light from the young stellar population. In addition, the studied region/galaxy can accrete mass from the outer regions and/or from the Intergalactic Medium (IGM) via infall. We define each one of these phases, and the processes they suffer from, in the following subsections.

\subsubsection{Infall}\label{Subsec:Infall}

Since we are not considering a closed-box model for our galaxy, it is necessary to take into account the interaction with its environment. This interaction can make the region either gain mass, via {\sl infall}, or lose mass due to galactic outflows.  As shown in \citet{gavilan13}, the observed properties of many galaxies can be reproduced by chemical evolution models without an outflow component, and therefore we have only included infall in the present tests in order to minimise the number of free parameters.  With this model, we will assess whether the basic relations involving the dust abundance ratios, as observed in the local Universe, may be qualitatively reproduced and understood in terms of the underlying physical processes.  More complicated models aimed to provide a detailed match to particular objects and/or other galaxy types will be explored in our future work.

Here we have considered, for the sake of simplicity, that the infall rate from the IGM can be approximately described by an exponential law:
\begin{equation}\label{Eq:Infall}
	\mathrm{I}(\mathrm{t})= \frac{\Sigma_\mathrm{total}}{\tau \ (1- e^{\frac{-\mathrm{T}_0}{\tau}})} \ e^{\frac{-\mathrm{t}}{\tau}},
\end{equation}
which depends on two parameters, the infall timescale, $\tau$, controlling the infall rate, and $\Sigma_\mathrm{total}$, the final total surface density of the region under study at $\rm t = T_{0}$. These quantities, $\Sigma_{tot}$ and $\tau$, which fully define the infall rate, are model inputs meaning that they must be changed to represent different galaxies or regions
\begin{figure}
\centering
\includegraphics[width=0.48\textwidth,angle=0]{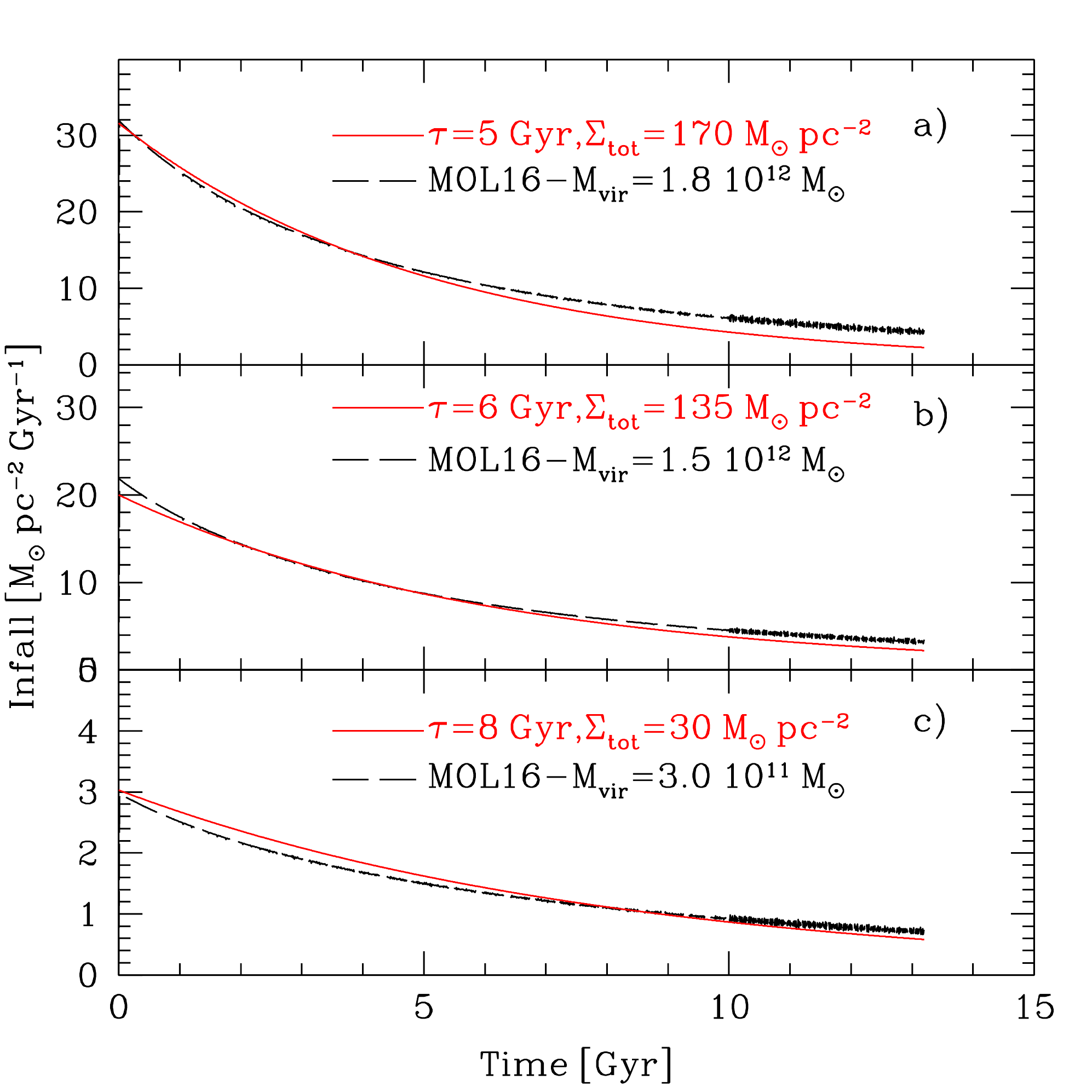}
\caption{Temporal evolution of the infall rate for three different
  combinations of final total surface density, $\Sigma_{tot}$, and collapse
  time scale, $\tau$, with red solid lines, as labelled in each panel,
  compared with similar infall rates as obtained in \citep{Molla_2016}
  for galaxies with different virial masses --dashed black lines.}
\label{Fig:infall}
\end{figure}

According to Eq.~\eqref{Eq:Infall}, if $ \tau \ll \mathrm{T}_0$, the majority of the gas will have already infallen on a rapid timescale. On the other hand, if $ \tau \ga \mathrm{T}_0$, the gas will fall at an approximately constant rate.  The above expression captures the main features of more realistic infall rates obtained in other models and cosmological simulations.

We show in Fig.~\ref{Fig:infall}, three different combinations of $\Sigma_{\mathrm{total}}$ and $\tau$, as labelled in each panel, compared with the infall rates used in \citet{Molla_2016} for galaxies of different virial masses/sizes.  We assume that this gas falling into the galaxy is ionised, since it comes directly from the IGM and/or from a hot galactic halo, where the temperature is well above $10^{4}$~K. Furthermore, the gas has primordial abundances, i.e., $\mathrm{X}_0 = 0.76$ and $\mathrm{Y}_0 = 0.24$, $Z=0$ \citep{coc12}. This assumption of primordial abundances may not be entirely correct for fountain material mixed with coronal material infalling back onto the disk as explored in \citet{gibson02}; however, it is probably sufficient as a first order hypothesis.

\subsubsection{Recombination}\label{Subsubsec:Recombination}

The gas in the ionised phase will capture free electrons and become neutral on a characteristic recombination timescale:
\begin{equation}
\tau_{\rm rec}(t) = \frac{1}{ n_{\rm e}(t)\, \langle\sigma v\rangle_{\rm B} },
\end{equation}
where $\rm n_{i}(t)$ denotes the electron number density at time $\rm
t$, and $\langle\sigma v\rangle_{\rm B} = 2.59 \times
10^{-13}$~cm$^{3}$~s$^{-1}$ corresponds to the thermally-averaged
Case-B recombination cross-section (i.e. without considering transitions to the ground state) appropriate for a temperature of the ionised phase of $10^4\,$K \citep{AGN2}.

\subsubsection{Molecular hydrogen formation}\label{Subsubsec:Molecular}

The conversion of atomic into molecular hydrogen is another relevant
physical process given the importance of the molecular phase in the
star formation process (see Section~\ref{Subsubsec:Star formation}).
There are several channels for the creation of H$_2$ molecules in the
diffuse ISM, such as:
\begin{eqnarray}\label{Eq:H2_nodust}
 \rm H^{-}+H & \rightarrow & H_2 + e^{-} \\
\rm H^{+}+H & \rightarrow & H^{+}_2 + \gamma   
\end{eqnarray}

Due to the low abundance of negative ions in the ISM, these reactions
have very slow rates.  Other channels, such as the three-body
reactions that take place in the Earth's atmosphere, are strongly
suppressed at astrophysical densities, and the most efficient
processes for the formation of $\rm H_2$ involve the interaction of H
atoms on the surface of dust grains, which are regulated by the
timescale:
\begin{equation}
\tau_{\rm cloud} = \frac{1}{2\,n_{\rm grains}\,R_{\rm H_2}},
\end{equation}
where $n_{\mathrm{grains}}$ is the volume density of dust grains in
the ISM, and $R_{\rm H_2}$ is the thermally-averaged cross-section for
the formation of molecular hydrogen \citep[see
  e.g.][]{H2_rate_Tielens_Hollenbach_1985, Draine_Bertoldi_96,
  H2_formation_in_grains}.

According to \citet{H2_formation_low_Z}, $\rm H_2$ formation in grains
becomes dominant as soon as $Z > Z_{\rm lim} \sim 10^{-5}$.  This
value is so low that a detailed treatment of other channels would have
a minimal impact on the results, but it would be necessary to include them
in the model in order to start the process, since the initial
abundance of metals and dust grains is zero, and stars only form from molecular clouds.  We circumvent this problem, \citep[as in ][]{Molla_2017}, by
adding the minimum metallicity $Z_{\rm lim}$ to the value $Z$ of the
ISM when estimating the timescale for cloud formation. This way:
\begin{equation}\label{Eq:H_2_formation_Draine_Bertoldi}
n_{\rm grains}\,R_{\rm H_2} \approx \frac{Z+Z_{lim}}{Z_{\sun}}\,n_{\rm ISM}\,R_{\rm H_2\sun}
\end{equation}
where $n_{\mathrm{ISM}}$ denotes the gas density, $\rm Z_{\sun}=
0.014$ is the solar metallicity \citep{asp09}, and $\rm R_{\rm
  H_2\sun} = 6.\,10^{-17}\,cm^{3}\,s^{-1}$ is the rate of molecular
hydrogen formation given by \citet{Draine_Bertoldi_96} for $\rm T =
100\,$K.  This is an approximation that could be eliminated in future
work by explicitly model the mechanisms responsible for the formation
of molecular clouds in the low-metallicity regime.

\subsubsection{Photoionisation and Photodissociation}
\label{Subsubsec:Ionising}

In order to estimate the photoionisation rate of H atoms at each time
step $t$, we compute the number of ionising photons produced in the
galaxy per unit time by summing the contribution:
\begin{equation}
Q_{\rm H}(Z',t') = \int_{0}^\text{912\,\AA} \lambda\, \frac{L_{\lambda}(Z',t')}{h\,c}\, \dd\lambda
\end{equation}
of different Single Stellar Populations (SSPs) of age $\rm t'$ and
metallicity $Z'=Z(t-t')$, where $L_{\lambda}(Z',t')$ is the specific
luminosity per unit wavelength $\lambda$ per unit mass of the SSP, $h$
is the Planck constant, and $c$ is the speed of light.  The total rate
of ionising photons at time $t$ is:
\begin{equation}\label{eq:Ionizing_photons}
R_\mathrm{ion}(t) = \int_0^t  Q_{\rm H}(t',Z') \ \Psi(t-t')\,\dd t'
\end{equation} 
may be obtained by integration over the star formation history
$\Psi(t-t')$ and chemical evolution $Z(t-t')$ of the galaxy.

Since most of the ionising radiation of a SSP comes from young blue
stars that die in the first $10-100$\,Myr, it is possible to
approximate $\Psi(t-t') \sim \Psi(t)$ and obtain a constant
photoionisation efficiency per unit Star Formation Rate (SFR):
\begin{equation}
\eta_\mathrm{ion} \equiv \frac{ R_\mathrm{ion}(t) }{ \Psi(t) } \simeq \int_0^t  Q_{\rm H}(t',Z')\,\dd t'
\ \equiv\ \eta_{\rm ion}\,\Psi(t)
\end{equation}
Taking the values of $Q_{\rm H}(Z',t')$ tabulated for the {\sc
  POPSTAR} code \citep{POPSTAR} for SSPs with ages between
$0.1\,\mathrm{Myr}$ and $15\, \mathrm{Gyr}$, we obtain $\eta_{\rm
  ion}=955.29$ for a \citet{IMF_Kroupa} IMF.

In order to obtain the efficiency of photodissociation of
$\mathrm{H}_2$ into $\rm HI$, we perform a similar computation for the
radiation emitted by each SSP in the Lyman-Werner band:
\begin{equation}
Q_{\rm LW}(Z',t') = \int_\text{912\,\AA}^\text{1107\,\AA} \lambda\, \frac{L_{\lambda}(Z',t')}{h\,c}\, \dd\lambda.
\end{equation}
However, \citet{Draine_Bertoldi_96} showed that dust grains may absorb
up to $\sim 60\%$ of the photons capable of dissociating hydrogen
molecules, and a large fraction of the remaining photons actually
excite different rotational and vibrational states, reducing their
dissociation probability to $\sim 15\%$.  Therefore, the net
photodissociation efficiency per unit SFR should be of the order:
\begin{equation}
\eta_{\rm diss} \equiv \frac{ R_\mathrm{diss}(t) }{ \Psi(t) } \sim 0.4 \times 0.15 \int_{0}^{t} Q_{\rm LW}(t')\ \dd t' = 380.93.
\end{equation}

From  the aforementioned considerations, we see that the
conversion of phases can be defined without the need for any free
parameters. 

\subsection{Stars}\label{Subsec:Stars}
	\subsubsection{Star formation}\label{Subsubsec:Star formation}

The birthrate, i.e., the amount of stars of each mass that have been
created per unit time and per unit area, is usually decomposed
\citep{pagelbook} as:	
\begin{equation}\label{Eq:Birthrate}
\rm B(\mathrm{t},m) = \Psi(\mathrm{t})\,\phi(\rm m),
\end{equation}
where $ \Psi(\mathrm{t})$ is the SFR, defined as the mass in
newly-born stars created per unit time, and $\rm \phi(m)$ is the
initial mass function (IMF).

Despite its crucial importance in galactic evolution, there is not a precise
theoretical model that takes into account all the effects associated with star formation. Thus, the use of empirical laws to define the
SFR is the traditional approach employed. Usually,
these laws relate the SFR with the local (two- or three-dimensional)
density of the gas.  One of the most common prescriptions is the
Kennicutt-Schmidt law based on \citet{SFR_Schmidt} and \citet{SFR_Kennicutt}.
The Kennicutt-Schmidt law is expressed by the well known formalism: $\rm
\Psi(\mathrm{t}) \propto \Sigma_{\mathrm{gas}}^n(\mathrm{t})$, where
$\Sigma_{\mathrm{gas}}$(t) is the surface density of gas and $n$ is an
exponent whose best-fitting value is $\rm n = 1.4 \pm 0.15$.

However, there have been studies suggesting that the correlation
between SFR and molecular gas surface density is stronger than the one
between SFR and total gas surface density, \citep[see
  e.g.][]{Wong_Blitz2002, Krumholz_Thompson2007,
  Robertson_Kravtsov2008, SFR_molecular_obs_Bigiel+08, Krumholz2009,
  SFR_molecular_obs_Bigiel+10, SFR_molecular_obs_Bolatto+,
  Schruba2011,Krumholz2012, SFR_molecular_obs_Leroy}, which is
reasonable since it is clear that stars form within cold molecular clouds.
Therefore, given that our model considers the multiphase structure of
the galactic region/galaxy, we prefer to use a relationship between
the SFR and the molecular gas density, such as: $\Psi(\mathrm{t}) =
\frac{\Sigma_{\mathrm{m}}(\mathrm{t})}{\tau_{\Psi}},$ where
$\Sigma_{\mathrm{m}}$ is the mass surface density of the molecular
phase and $\tau_{\Psi}$ is a characteristic timescale of star formation. 

There are uncertainties in the value of $\tau_{\Psi}$ and its possible
variability, depending on the medium where star formation is
happening. On the observational side,
\citet{SFR_molecular_obs_Bigiel+08} studied 7 spiral and 11
late-type/dwarf galaxies and found a mean value of
$<\tau_{\Psi}>=2.0$\,Gyr, detecting differences between the centre and
the outskirts. In fact, \citet{SFR_molecular_obs_Bigiel+10} observed
later the outer part of the disks of spiral galaxies and concluded
that $\tau_{\Psi}$ is even higher than the Hubble time in these outer
regions; \citep{SFR_molecular_obs_Bolatto+} estimated a slightly
shorter $<\tau_{\Psi}> = 1.6$\,Gyr,  with significant variations (ranging form 0.6 to 7.5\,Gyr) throughout the
Small Magellanic Cloud. Finally, \citet{SFR_molecular_obs_Leroy} have
obtained a value $\tau_{\Psi}= 2.2\pm 0.3$\,Gyr from observations of 30 nearby
disk galaxies. Consequently, there have been some attempts to make empirical
laws considering different depletion times depending on the
environment, such as \citet{Krumholz2009}, whose empirical law depends
on the surface density of the ISM where the star formation occurs.  We
are going to use here the expression given by these last authors:
\begin{equation}
    \Psi(\mathrm{t}) = \frac{\Sigma_{\mathrm{m}}}{2.6 \ \mathrm{Gyr}} 
    \begin{cases}
    \left(\frac{\Sigma_{\mathrm{ISM}}}{85.0 \mathrm{M_{\odot}\,pc^{-2}}} \right)^{-0.33} \mathrm{if} \ \Sigma_{\mathrm{ISM}}< 85\, \mathrm{M_{\odot}\,pc^{-2}},\\ 
    \left(\frac{\Sigma_{\mathrm{ISM}}}{85.0 \mathrm{M_{\odot}\,pc^{-2}}} \right)^{0.33} \ \mathrm{if} \ \Sigma_{\mathrm{ISM}}> 85\, \mathrm{M_{\odot}\,pc^{-2}}
  \end{cases}
\end{equation}

The IMF is the observational distribution of the number of stars
according to their initial masses in every star formation event. There
are also some uncertainties in the exact shape of the IMF, particularly
in the low and high mass regime due to the difficulties of observing
stars with extremely low masses and to the rapid death of massive stars. In this work, we have assumed an IMF from \citet{IMF_Kroupa},
with the following expression:
\begin{equation}\label{IMF}
\rm  \phi(m) = 
  \begin{cases} 
   \xi_1 \mathrm{m}^{-1.3} & \text{if }  0.1\,\mathrm{M_{\odot}} <  \mathrm{
   m} < 0.5\,\mathrm{M_{\odot}}   \\ 
   \xi_2 \mathrm{m}^{-2.3} & \text{if } \mathrm{m}> 0.5\,\mathrm{M_{\odot}},
  \end{cases}
\end{equation}
where $\xi_1 = 0.4690$ and $\xi_2 = 0.2345$ are the normalisation
constants, for a minimum stellar mass of $ \rm m_{\rm low} = 0.1\,
M_{\odot}$ and a maximum stellar mass of $\rm m_{\rm up} = 50
\,M_{\odot}$, which give $\rm \int_{m_{low}}^{m_{up}}m\,\phi(m)dm=1$ .
Our upper limit was chosen in order to minimise the extrapolation of \citet{Ejecta_SN_CC} ejecta, which are provided up to 40~M$_\odot$, and it agrees with the upper limit found or used by other authors \citep[e.g.][]{Kobayashi2011, Mishurov2018, Mishurov2019}.
\footnote{Furthermore, for the adopted IMF, the contribution of stars above this limit to the total ejecta budget is negligible. See section \ref{Subsubsec:Ejecta}.}

\subsubsection{Mean stellar lifetimes}\label{Subsubsec:Lifetime}

Instead of using the instantaneous recycling approximation, we will
take into account the lifetimes of the stars with the empirical
formulae provided by \citet{Lifetimes}, calculated from different
stellar tracks \citep{Lifetimes_Data_1,Lifetimes_Data_2,
Lifetimes_Data_3} over a stellar mass range of 
[$0.6 \ \mathrm{M}_{\odot}$ - $120 \ \mathrm{M}_{\odot}$] and a
metallicity range of [$0.0004$ - $0.05$]:
\begin{equation} \label{Eq:lifetime}
\rm \log{\tau_{*}(m,Z)} =  a_0(Z) + a_1(Z) \log{m} + a_2(Z)\,(\log{m})^2,
\end{equation}
where $\tau_{*}(m,Z)$ denotes the lifetime of a star of mass $m$ and
metallicity $Z$, and the coefficients $a_{0}(Z)$, $a_{1}(Z)$ and
$a_{2}(Z)$ are given by the following expressions:
\begin{eqnarray}
\nonumber
\rm  a_{0}(Z) &=&  10.13 + 0.07547 \log{\rm Z} - 0.008084 (\log{\rm Z})^{2} \\
\nonumber  
 \rm a_{1}(Z) &=& -4.424 - 0.7939 \log{\rm Z} - 0.1187   (\log{\rm Z})^{2} \\
  \nonumber
 \rm a_2(Z)   &=&  1.262 + 0.3385 \log{\rm Z} + 0.05417  (\log{\rm Z})^{2}
\end{eqnarray}

\subsubsection{Stellar ejecta}\label{Subsubsec:Ejecta}

The chemical composition of the gas ejected by a star at its death is
different from the one of the molecular cloud from which it formed, as
the material has been enriched with the elements that have been
created in the interior of the star via nucleosynthesis.  For our
chemical evolution model, we are interested in the total (gas and
dust) ejection rate $\rm E_{j}(t)$ of each element $j$ as a function
of time $t$, which we divide in three components:
\begin{equation}
	\mathrm{E}_{j} (\mathrm{t})= \mathrm{E}_{j,{\mathrm{AGB}
        }}(\mathrm{t}) + \mathrm{E}_{j,{\mathrm{SNcc} }}(\mathrm{t}) +
        \mathrm{E}_{j,{\mathrm{SNIa}}}(\mathrm{t}),
\end{equation}
where $\mathrm{E}_{j,{\mathrm{AGB} }}(\mathrm{t})$,
$\mathrm{E}_{j,{\mathrm{SNcc} }}(\mathrm{t})$ and
$\mathrm{E}_{j,{\mathrm{SNIa}}}(\mathrm{t})$ denote the contributions
of low- and intermediate-mass stars, high mass stars and type Ia SNe,
respectively.

Low and intermediate mass stars end their lives as planetary nebulae
after passing the Asymptotic Giant Branch (AGB) phase, during which
the star loses an important fraction of its mass. These objects eject
to the ISM mainly $\rm ^{12}C$, $\rm ^{13}C$, $\rm ^{14}N$ and $\rm
^{19}F$. The ejection rate for each element is computed as:
\begin{small}
    \begin{equation}
   \rm \mathrm{E}_{j,\mathrm{AGB}}(\mathrm{t}) = \int_{m_{*,\rm lim \rm (t)}}^{8\,M_{\sun}} \Psi(\mathrm{t'})\,\phi(m)\,e_{j,{\rm AGB}}(m,Z(\mathrm{t}'))\,dm,
    \label{ejAGB}
    \end{equation}
\end{small}
where $m_{\rm lim}(t,Z)$ is the minimum mass of a star of metallicity
Z to have died prior to the moment $\mathrm{t}$.  These stars are born
at a time $\mathrm{t}'=\mathrm{t}-\tau_{*}(m,Z)$, according to
equation~(\ref{Eq:lifetime}) above.  $\Psi(\mathrm{t}')$ is the SFR of
the galaxy/region at that time, and $\phi(m)$ is the assumed IMF.  For 
the mass $e_{j,{\rm AGB}}(m,Z(\mathrm{t}'))$ returned by a low and intermediate mass star with mass $m$ and metallicity $Z$ at the end of its life in form of element
$j$, we use the tables provided by \citet{AGB_gas_ejecta}.

The final fate of high mass stars ($m> 8\,M_{\sun}$) is that of an explosion
as core-collapse supernova. These supernovae are responsible for
ejecting the so-called $\alpha$-elements (i.e., $^{12}$C, $^{16}$O,
$^{20}$Ne, $^{24}$Mg, $^{28}$Si, $^{32}$S and $^{40}$Ca) created in
the interior of these massive stars during their evolution, as well as
unstable isotopes of Ti, Cr, Fe and Ni, created during the explosion
itself, whose decay provides a significant fraction of the energy that
is radiated away during the explosion.

At any given time, the rate of core-collapse supernova explosions is given by:
\begin{equation}
	\mathrm{R}_{{\rm SNcc}} (\mathrm{t}) = \int_{8\, \mathrm{M_{\sun}}}^{m_{\rm up}}\,\Psi(\mathrm{t}')\,\phi(m)\,dm
\end{equation}
and the ejection rate of element $j$ is simply
\begin{equation}
	\rm \mathrm{E}_{j,\mathrm{SNcc}} (\mathrm{t}) = \int_{8\,\mathrm{M_{\sun}}}^{m_{\mathrm{up}}}  \Psi(\mathrm{t}')\, \phi(m)\,e_{j,{\mathrm{SNcc} }} (m,Z(\mathrm{t}'))\,dm,
\end{equation}

where $e_{j,{\mathrm{SNcc} }}(m,Z)$ is the total mass
that a massive star of mass $m$ and metallicity $Z$ returns as element $j$ at
the end of its life taken from \citet{Ejecta_SN_CC}\footnote{These tables have been extensively used in the literature, and there is little difference in the oxygen yields up to $m_{up}=50\,M_{\odot}$ with respect to more recent prescriptions \citep{AGB_gas_ejecta}.}.

Finally, the third mechanism that ejects newly synthesised metals to
the ISM is the thermonuclear explosion of type-Ia supernovae. Although
some other chemical species are produced, their main contribution is
the synthesis of iron-peak elements (Ti, V, Cr, Mn, Fe, Co and Ni),
being the principal mechanism for iron enrichment in galaxies.

In order to compute the ejecta of type-Ia supernovae, we need their
rate of occurrence:
\begin{equation}
\rm \mathrm{R}_{{\rm SNIa}} = k_{\alpha} \int_{\tau_{\rm min}}^{min(\mathrm{t},\tau_{\rm max})}
\!\!\!\! A_B(\mathrm{t}-\tau) \ \Psi(\mathrm{t}-\tau) \ \mathrm{DTD}(\tau)\, d\tau,
\end{equation}
where $k_{\alpha}=\int_{m_{\rm low}}^{m_{\rm up}}\,\phi(m)$\,dm is the
number of stars per unit mass in a stellar generation, $\rm A_B$ is
the fraction of (binary) systems that end up as a thermonuclear SN,
and DTD is the Delay Time Distribution that describes how many type Ia
SNe progenitors die after a at time $\tau$.  The lower limit of the
integral is the minimum delay time $\tau_{\rm min}$, whereas the upper
limit is the minimum between the maximum delay time and the current
age $t$ of the galaxy.  We use a constant $\rm A_B= 0.10$ for the
Kroupa IMF \citep{Maoz_Hallakoun_2017} and the empirical DTD from \citet{DTD_Maoz}.
The precise shape of the DTD is a very active research topic, with some slight differences of the empirical functions with respect to theoretical predictions \citep[e.g.][]{Binary_fraction}. These latter properly account for stellar lifetimes, binary fractions, and evolution, while the first ones are easy to use, but not totally correct at the shortest phases of the evolution ($t< 80$\,Myr). However, since we are not computed relative abundances of different chemical species, as $\alpha/Fe$, but only the whole metallicity Z, it is unlikely that these differences would produce a significant effect in the present context.

Once the supernova rate is determined, the ejection rate of type Ia supernovae is:
\begin{equation}
\rm 	\mathrm{E}_{j,{\mathrm{SNIa} }} (\mathrm{t})= \mathrm{R}_{{\mathrm{SNIa}}}\,e_{j, \mathrm{Ia}},
\end{equation}
where $e_{j,\mathrm{Ia}}$ are the ejecta of element $j$ in each SNIa, for which we adopt the yields of the classical model W7 from \citet{Ejecta_SN_Ia}.

With these three definitions, the amount of mass per unit time that
stars inject into the ISM in the form of hydrogen and helium will be
denoted by $E_\mathrm{X}(t)$ and $E_\mathrm{Y}(t)$, respectively.  As
will be discussed below, the total ejection rate of metals:
\begin{equation}
E_{\rm Z_{total}}(t) =
\sum_j \left[\,E_{j,\mathrm{AGB}}(t) + E_{j,\mathrm{SNcc}}(t) + E_{j,\mathrm{SNIa}}(t)\,\right]
\end{equation}
will distinguish between the gaseous and dust phase, $E_{\rm Z_{total}}(t) = E_{\rm Z_{gas}}(t) + E_{\rm dust}(t)$.

\subsection{Dust}\label{Subsec:Dust}

Given the relevance of the dust in some key processes of galactic
evolution, it is important to model correctly its life cycle. As noted previously, dust grains are mainly created
in the AGB phase of low and intermediate mass stars ($m<8
\mathrm{M_{\sun}}$) and at the final stage of the shock waves produced
by core-collapse and thermonuclear SNe.  Several authors
\citep[e.g. Z08, ][]{Zhukovska2016, Accretion_McKinnon,
  Accretion_Gioannini, Accretion_Aoyama, Zhukovska2018} have argued
that in order to reproduce the observed values of the dust to gas
ratio, it is necessary that dust grains accrete mass from the
ISM\footnote{In addition to the dust growth by accretion, dust grains
  suffer a variety of effects in the ISM that change their size, such
  as shattering, coagulation, etc.  They are not considered here,
  since we assume an unchanging \citet{MRN}'s grain size
  distribution.}.  Finally, we also consider that the grains may be
destroyed by SN shock waves, thermal sputtering, radiative torques and
astration\footnote{Mass loss of certain chemical element or dust
  species due to its incorporation to a star during a SF process}.

\subsubsection{Dust creation}\label{Subsubsec:Dust_creation}

The creation of dust grains inside the atmosphere of low- or
intermediate-mass stars is not significant until they reach the AGB
phase \citep{Gail09}.  Afterwards, they are considered to be the
main stellar contributors to dust enrichment in the ISM.  At any given
moment $t$, the total mass of dust ejected per unit time is given
by:
\begin{equation}
E_{\rm d, AGB}(t) =
\int_{m_{\mathrm{lim}}(t)}^{8 \ \mathrm{M}_{\odot}} \!\!\!\!\!\!\!
    \phi(m)\, \Psi(\mathrm{t}')\ e_{\mathrm{d,AGB}}(m,Z_{\rm ISM}(\mathrm{t}'))
    \, dm,
\end{equation}

where $e_{d, \rm AGB}(m, Z_{\rm ISM}(\mathrm{t}'))$ is the dust ejecta for a star of mass $m$ and metallicity $Z_{\rm ISM}$. Yields are taken from \citet{AGB_Z_10_-3, AGB_Z_8_10_-3, AGB_Z_3_10_-4, AGB_Z_1.8_10_-2}.

Secondly, we have to consider the dust produced by the massive stars which end up their lives as core collapse SNe. In an equivalent equation:
\begin{equation}
E_{\rm d, SN cc}(t) =
\int_{8\,\mathrm{m}_{\odot}}^{m_{\rm up}} \phi(m)\, \Psi(t')\ e_{\rm d, SNcc}(m,Z(t'))\,dm,
\end{equation}

where $e_{\rm d,SNcc}(m,Z(t))$ is the mass of dust injected to the ISM
when a star of mass $m$ and metallicity $Z$ explodes as a core
collapse SNe.  Even though the SN shock waves are the main mechanism
of dust destruction, there is evidence of dust in the surroundings
of supernova remnants \citep{Dust_SN1987A}.  It is extremely difficult
to measure the mass of the progenitor, making it all but impossible to obtain an
empirical value of these ejecta. Hence, we use the results obtained by
\citet{SN_cc_dust} from simulations of dust grains formation and their
survival to the reverse shock wave in SNe. We adopt the ejecta
corresponding to $\rm n_{\mathrm{ISM}} = 1\,cm^{-3}$ as a
representative value of the mean density of the ISM.

The last stellar contributors to dust formation in a galaxy are the type Ia SNe:
\begin{equation}
E_{\rm d, SNIa}(t) = R_{\rm SNIa}\, e_{\rm d, SNIa},
\end{equation}
where $e_{\rm d,SNIa}$, is the ejecta of dust for each SN-Ia.  In the
current work we use the data of \citet{Nozawa2011}, having
simulated dust production using the carbon-deflagration W7 model.  These
authors argue that the reverse-shock that is created in type Ia SNe
destroys the majority of the created dust mass. Thus, we assume that only a small amount of dust ($\rm 0.01~M_{\odot}$,
for a density of the environment near the explosion of $ n_{\rm e} =
0.01 \ cm^{-3}$) survives the reverse shock. Broadly speaking, 
type Ia SNe are minority contributions to dust production, but we account for them, regardledd, to maintain consistency between the gas and dust ejecta.

Summing all contributions, the total ejected dust mass per unit time is:
\begin{equation}
E_{\rm dust}(t) = E_{\rm d,AGB}(t) + E_{\rm d,SNcc}(t) + E_{\rm d,SNIa}(t).
\end{equation}

We must now take into account that dust particles are composed of solid metals.
In order to model in a self-consistent way the yields of the gas phase, $E_{\rm Z_{gas}}(t) = E_{\rm Z_{total}}(t) - E_{\rm dust}(t)$, we subtract the amount of the chemical elements that are depleted in each dust grain species from the total ejecta.

\subsubsection{Dust growth}\label{Subsubsec:Dust_accretion}

As noted in Section \ref{sec:Intro}, 
evidence exists to suggest that dust growth by accretion of gas inside giant molecular clouds (GMCs) is necessary, to obtain the DTG values observed in nearby
galaxies.

From an observational perspective, according to \citet{Remy14} and
\citet{De_Vis2017}, the accretion timescale inside molecular clouds
should be very small, $\tau_g \approx 5\,\mathrm{Myr}$, in order to
explain the observed DTG in nearby galaxies. The former argue that
core collapse SNe are not able to contribute much to the dust budget
of the galaxy and, therefore, a very small growth is needed to explain
the DTG-Z trend. However, given that the uncertainties in the
determination of molecular gas and dust are still large, it is better
to use prescriptions that only depend on the properties of the
environment where the grains grow and not a time scale obtained from
GMC studies.

From a theoretical perspective, \citet{Hirashita2011} have developed a
model that takes into account the effects of the environmental
conditions where the dust grows, such as metallicity and number
density of the molecular clouds and other physical properties of the
grains, including sticking coefficient, size and temperature, obtaining the following
timescale for the dust grains growth:
\begin{footnotesize}
    \begin{equation}\label{Eq:Accretion}
	    \rm \tau_g = 6\times10^{-2} \left(  \frac{a}{0.1 \mu m}\right) \left(  \frac{Z_{\odot}}{Z}\right) \left(  \frac{1000 \  \mathrm{cm}^{-3}}{n_{\mathrm{cloud}}}\right) \sqrt{\frac{50}{T}} \left(\frac{0.3}{S}\right)\,[\mathrm{Gyr}],
\end{equation}
\end{footnotesize}
where $<a>$ is the mean radius of the MRN distribution, assumed to be
$<a>=0.08\,\mu m$, $Z_{\odot}=0.014$ is the solar metallicity, $\rm T$
is the temperature of the atomic phase, taken as 100\,K,
$n_{\mathrm{cloud}}$ is the number density of the molecular clouds and
$S$ is the sticking coefficient, defined as the probability
of an atom to {\it stick} when it collides with a dust grain. These last two coefficients are not entirely known {\it a priori} and, therefore, are deemed
free parameters.  Once computed, the growth in dust mass in the ISM \citep{Hirashita2011,
  hirashita19} due to accretion is:
\begin{equation}
  \left(\frac{d \Sigma_d}{d t}\right)_{\mathrm{acc}}= \left(1- \frac{\Sigma_d}{\Sigma_Z}\right)\frac{\Sigma_d}{\tau_{g}}f_{m}.
\end{equation}

\subsubsection{Dust destruction}\label{Subsubsec:Dust_destruction}

The other relevant processes involved in the life cycle of the dust are the ones that destroy grains. These processes are:
sputtering, collision with cosmic rays, supernova shock waves,
astration, radiative torque of a powerful and anisotropic radiation
field,\ldots. In our model, we consider together all dust
grain destruction processes, including thermal sputtering, supernova shock waves
and radiative torques. We consider separately the dust mass loss due
to astration, that is, the mass loss of a specific element in the dust
during the star formation process, which is, obviously, proportional
to the SFR:

\begin{equation}\label{Eq:Dust_destruction astration}
  \rm  \left(\frac{d \Sigma_d}{d t}\right)_{\mathrm{ast}} =  -\frac{\Sigma_d}{\Sigma_{\mathrm{ISM}}} \Psi(t)
\end{equation}

The other destruction mechanisms are treated as a whole, considering
that there is a characteristic timescale of dust grain destruction in
the ISM. Again, since we do not have sufficient information regarding these
processes, we assume this timescale as a free parameter, which,
following \citet{Jones1994,Dust_survival_optimistic} varies in the
range $\rm \tau_{dest}=0.4-2.0$\,Gyr.  Thus, the destruction rate due
to SN shock waves, thermal sputtering and radiative torque is:

\begin{equation}\label{Eq:Dust_destruction_timescale}
       \rm \left(\frac{d \Sigma_d}{d t}\right)_{\mathrm{SNe, others}} = - \frac{\Sigma_d}{\tau_{dest}}
\end{equation}

From the above, we see that we only need to invoke three free parameters related
with the dust growth and destruction: the sticking value $\rm $, the
density of GMC clouds, $\rm n_{cloud}$, and the timescale for the
destruction $\tau_{\rm dest}$.

\subsection{Equation System}

\begin{figure}
\includegraphics[width=8.35cm]{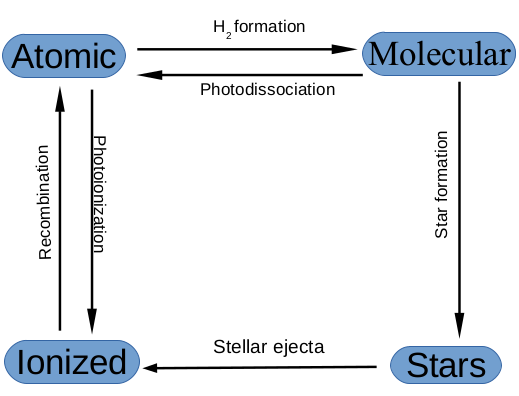}
\caption{Scheme of the multiphase structure of the ISM of our model
  and the processes of mass interchange among them.}	
\label{Fig:Phases}
\end{figure}

Considering all the processes that have been stated previously, the system of differential equations that govern the evolution of all the phases is:
\begin{eqnarray}
\dot\Sigma_\mathrm{a} &=&
  \frac{\Sigma_\mathrm{i}}{\tau_\mathrm{rec}} - \eta_\mathrm{ion} \Psi(t)
  - \frac{\Sigma_\mathrm{a}}{\tau_\mathrm{cloud}}  
  + \eta_\mathrm{diss} \Psi(t) \\
\dot\Sigma_\mathrm{i} &=&
  - \frac{\Sigma_\mathrm{i}}{\tau_\mathrm{rec}}
  + E_\mathrm{gas}(t) + \eta_\mathrm{ion} \Psi(t) + I(t) \\
\dot\Sigma_\mathrm{m} &=&
  \frac{\Sigma_\mathrm{a}}{\tau_\mathrm{cloud}} 
    - \Psi(t) - \eta_\mathrm{diss} \Psi(t) \\
\dot\Sigma_* &=& \Psi(t) - E_\mathrm{gas}(t) - E_\mathrm{dust}(t),
\end{eqnarray}
where $E_\mathrm{gas}(t) = E_\mathrm{X}(t) + E_\mathrm{Y}(t) +
E_\mathrm{Z_{gas}}(t)$ is the total ejected gas in the moment $t$,
i.e., the sum of the ejected gas of hydrogen, helium and metals.

This scenario of phases and processes is represented in the scheme of
Figure~\ref{Fig:Phases}.

We assume that all the ISM phases have identical chemical composition,
and we track the time evolution of the mass surface density of the
hydrogen (X) and helium (Y), as well as metallic elements in the gas
phase (Z) and in form of solid grains (dust):
 \begin{footnotesize}
\begin{eqnarray*}
   \dot{\Sigma}_{\rm X} &=& - \frac{\Sigma_{X}}{\Sigma_{\mathrm{ISM}}} \Psi(\mathrm{t}) + \rm E_{X} (\mathrm{t}) + \rm X_0 I_i(\mathrm{t}) \\
 \rm \dot{\Sigma}_{\rm Y} &=& - \frac{\Sigma_{\rm Y}}{\Sigma_{\mathrm{ISM}}} \Psi(\mathrm{t}) + \rm E_{Y} (\mathrm{t}) +\rm Y_0 \rm I_i(\mathrm{t}) \\
  \rm \dot{\Sigma}_{Z} &=& - \frac{\rm \Sigma_{Z}}{\rm \Sigma_{\mathrm{ISM} }} \Psi(\mathrm{t}) + \rm E_{Z_{gas}} (\mathrm{t}) + \frac{\Sigma_{\mathrm{d} }}{\tau_{\mathrm{dest}}} -  \left( 1 - \frac{\rm \Sigma_d}{\rm \Sigma_Z}\right) \frac{\Sigma_d}{\tau_{\mathrm{g}}} f_{\rm m}\\
\dot\Sigma_\mathrm{dust} &=& -\frac{\Sigma_\mathrm{dust}}{\Sigma_{\mathrm{ISM}}} \Psi(t) + \rm E_{dust}(\mathrm{t}) - \frac{\rm \Sigma_{d}}{\tau_{\mathrm{dest}}} + \left( 1 - \frac{\rm \Sigma_d}{\rm \Sigma_Z}\right) \frac{\rm \Sigma_d}{\rm \tau_{\mathrm{g}}} \rm f_{\rm m},
\end{eqnarray*}
\end{footnotesize}
where all terms and symbols have the meaning already given in previous
subsections and Table~\ref{tab_components}.

Due to the incompleteness and inconsistencies that arise when we
compare the tables from different dust sources, it is not very precise
to distinguish between the different dust species.  Thus, we have just
combined the ejecta of all the dust species from each source in order
to minimise artificial errors.  Furthermore, the extragalactic
observations from \citet{Vilchez+2019}, against which we will compare our models, do not distinguish among different species.  Therefore,
we are not discerning different dust species, considering the
evolution of all the metallic elements in dust grains at once.

\section{Computed Models and Results}
\label{sec:Results}

We have run 1350 models, varying the inputs defining different galaxy
sizes by $\Sigma_{\rm total}$, and $\tau$, and the three free
parameters related with the dust processes in the range shown in
Table~\ref{tab_run_models}.  From this set of models, we first analyse
the time evolution of a reference model assumed to be a MWG-type
galaxy. Then, we study the evolution of the dust to gas (DTG) and dust
to stars (DTS) ratios with the metallicity, for a subsample of models,
in order to examine the role that each free parameter plays in shaping the relationships for this reference model. Finally, we will compare the results obtained at the
end of the evolution, for the entire set of models, with the available
sets of data for two large samples of galaxies.

\begin{table}
\caption{Range of dust and infall parameters of our computed models.}
\label{tab_run_models}
\begin{center}
\begin{tabular}{ll}
\hline
Dust parameters & range \\
\hline
S & $[0.3,0.9]$ \\
$\tau_{\rm dest}$ & $[0.8,1.8]$\,Gyr \\
$\mathrm{n}_{\rm cloud}$ & $[1000,10000]\,\mathrm{cm}^{-3}$ \\
\hline 
Infall parameters & range\\
\hline
$\tau$ & $[0.1,20.0]$\,Gyr \\
$\Sigma_{\rm total}$ & $ $[10.0,200.0]$\,\frac{M_{\odot}}{\mathrm{pc}^{2}}$ \\
\hline
			\end{tabular}
	\end{center}
\end{table}

\subsection{Time evolution of MWG-type galaxy model}\label{subsec:Time_evolution}

\begin{table}
\begin{center}
\caption{Summary of observational constraints to compare with the MWG model}
\label{tab_obs}
\begin{tabular}{lll}
\hline
Data & Value  & Ref.\\
\hline
$\Sigma{\mathrm{total}}$ & 55 $\pm$ 10 \,$\rm M_{\sun}\,pc^{-2}$ & M15 \& BO17\\
$\Sigma_{\mathrm{*}}$ & 38.5 $\pm $ 2.5\,$\rm M_{\sun}\,pc^{-2}$ & BO17\\
$\rm \Sigma_{SFR}$ & 5.3 $\pm $  4.0\,$\rm M_{\sun}\,pc^{-2} Gyr^{-1}$&  M15\\
$\mu$ & 0.15 $\pm$ 0.08 & M15\\ 
$\rm \Sigma_{R_{CC}}$ &  0.0208 $\pm$ 0.0118 \,$\rm pc^{-2} Gyr^{-1}$ & T94 \\
$\rm \Sigma_{R_{Ia}}$ &  0.0062 $\pm $ 0.0047 \,$\rm pc^{-2} Gyr^{-1}$ & T94 \\
$\rm f_{m}$ & 0.4 $\pm$ 0.2 & M15\\
$\rm DTG$ & 0.0106 $\pm$ 0.007 & F03 \\
$\rm Z_{\sun}$ & 0.014 $\pm$ 0.008 & ASP09\\
$\rm Z_{gas}$ &  0.011$\pm$ 0.007 &  EST19\\

\hline
\end{tabular}
\end{center}
\footnotesize{References are: \citet{Tammann1994}(T94); \citet{Frisch2003}(F03); \citet{asp09}(ASP09); \citet{AGB_gas_ejecta}(M15), \citet{Bovy2017}(BO17) and \citet{est19}(EST19).}
\end{table}
We start by applying the model to a MWG-type galaxy, in order to provide a baseline calibration.  The characteristics used as observational
constraints are shown in Table~\ref{tab_obs}.  They refer to the
stellar density, $\Sigma_{*}$, (which would be similar for the
brighter galaxies, but it would be smaller for low brightness
galaxies), the metals abundance, $Z$, the fraction of gas,
$\mu=\Sigma_{gas}/\Sigma_{\rm total}$, the fraction of the molecular
gas to the total gas, $\rm f_{m}=\frac{\Sigma_{m}}{\Sigma_{gas}}$, the
dust to gas ratio, $\rm DTG=\frac{\Sigma_{d}}{\Sigma_{gas}}$, and the
core collapse and type Ia supernova rates.

As noted in Section~\ref{Subsec:Infall}, the assumed infall rate values,
$\Sigma_{\rm total}$ and $\tau$, are related with the final size of
the galaxy and we need to select them now to reproduce the MWG.  We
use $\tau=2.5$\,Gyr and $\Sigma_{\rm total}=55\,\rm M_{\sun}\,pc^{-2}$
for this reference model, consistent with the total density specified
in Table\,\ref{tab_obs}. Other different input parameters, $\rm
\Sigma_{total}$, and $\tau$ will be useful to reproduce galaxies of
different sizes and types, and will be explored in \S3.2. The MWG reference model
has been computed with the following free parameters: $\mathrm{n}_{\rm
  cloud} = 2500\,\mathrm{cm}^{-3}$, $\tau_{\rm dest} =
1.3\,\mathrm{Gyr}$ and $\rm S = 0.7$.

\begin{figure*}
\centering
\includegraphics[width=0.49\textwidth,angle=0]{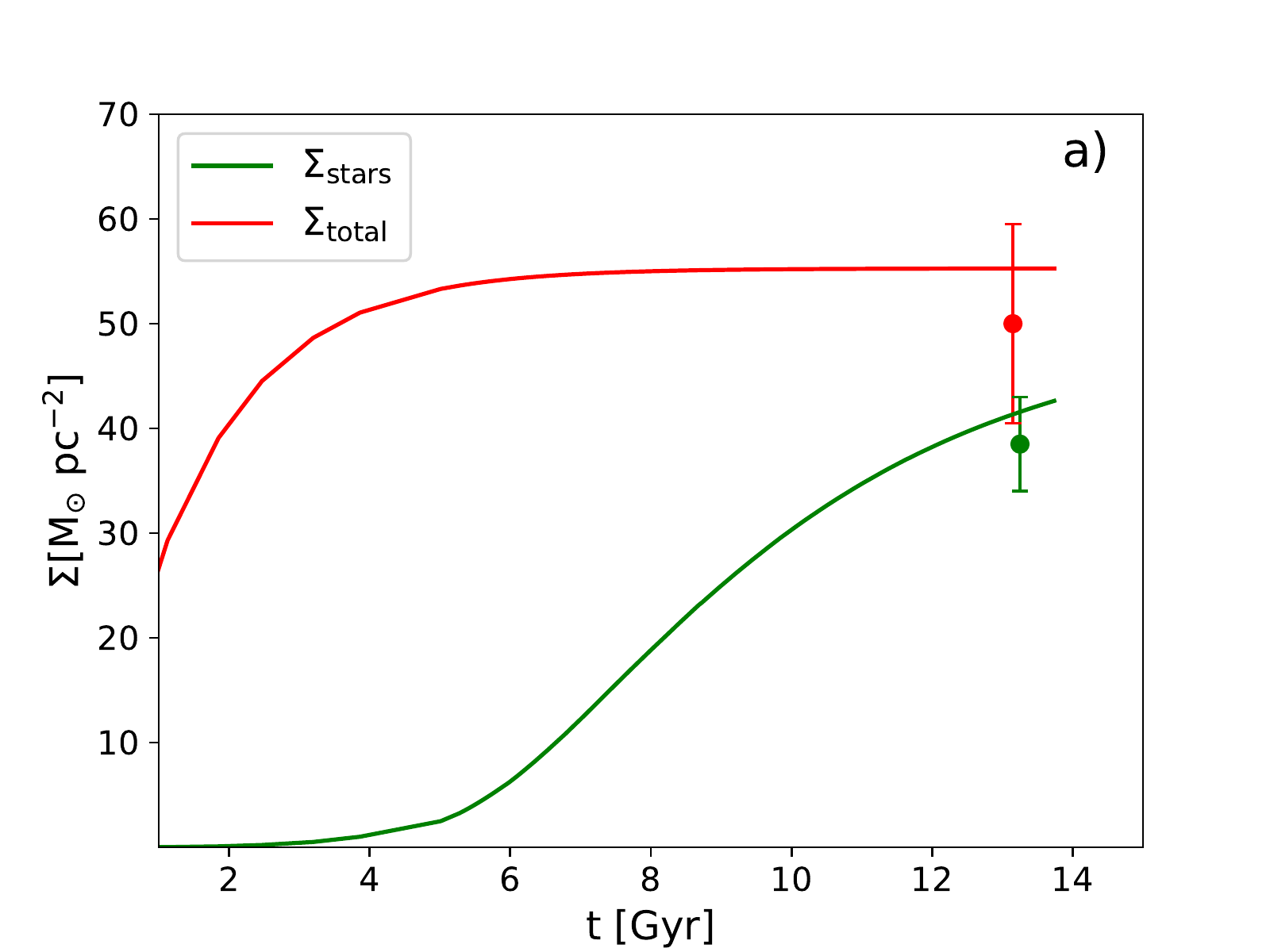}
\includegraphics[width=0.49\textwidth,angle=0]{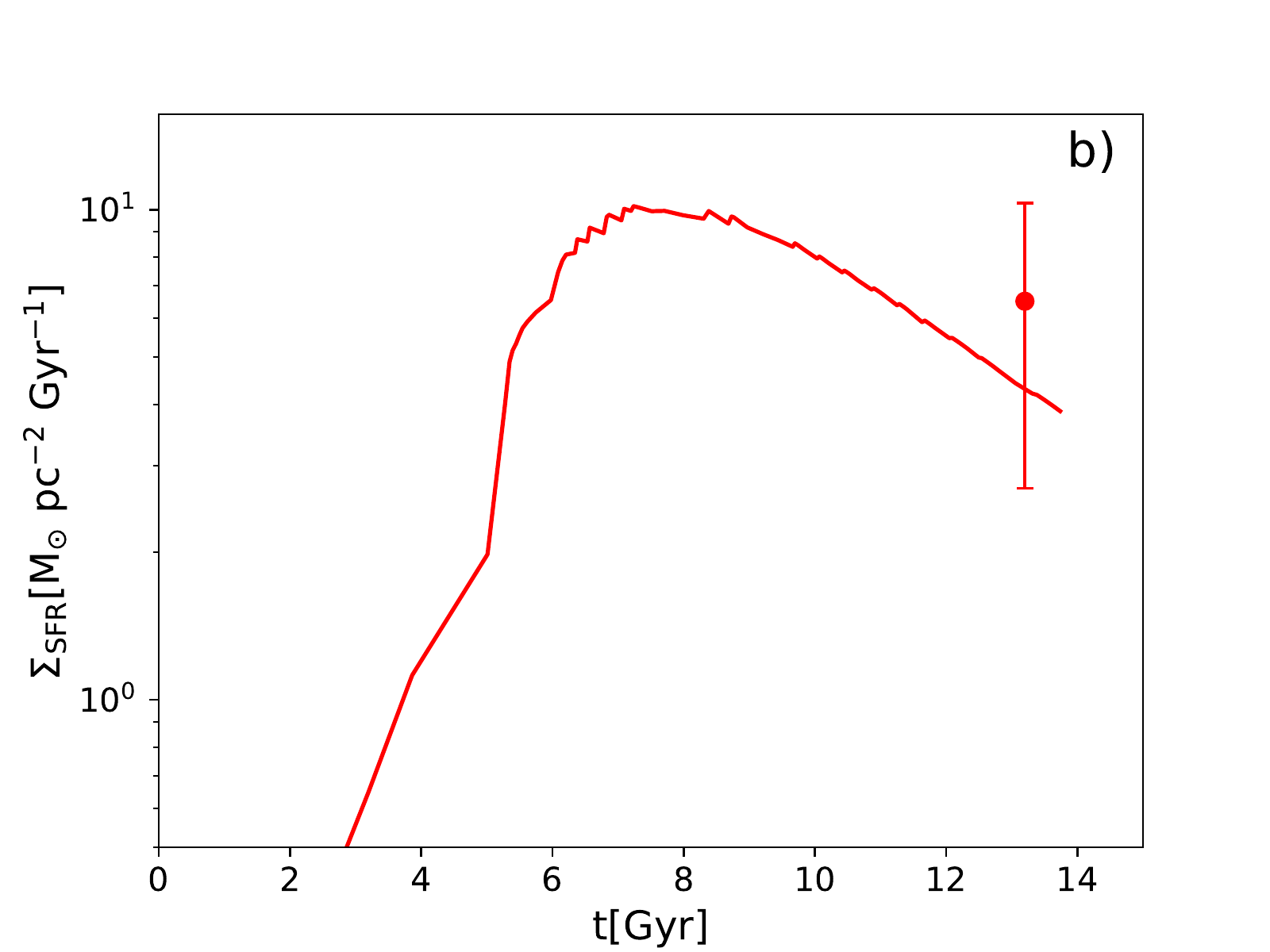}
\includegraphics[width=0.49\textwidth,angle=0]{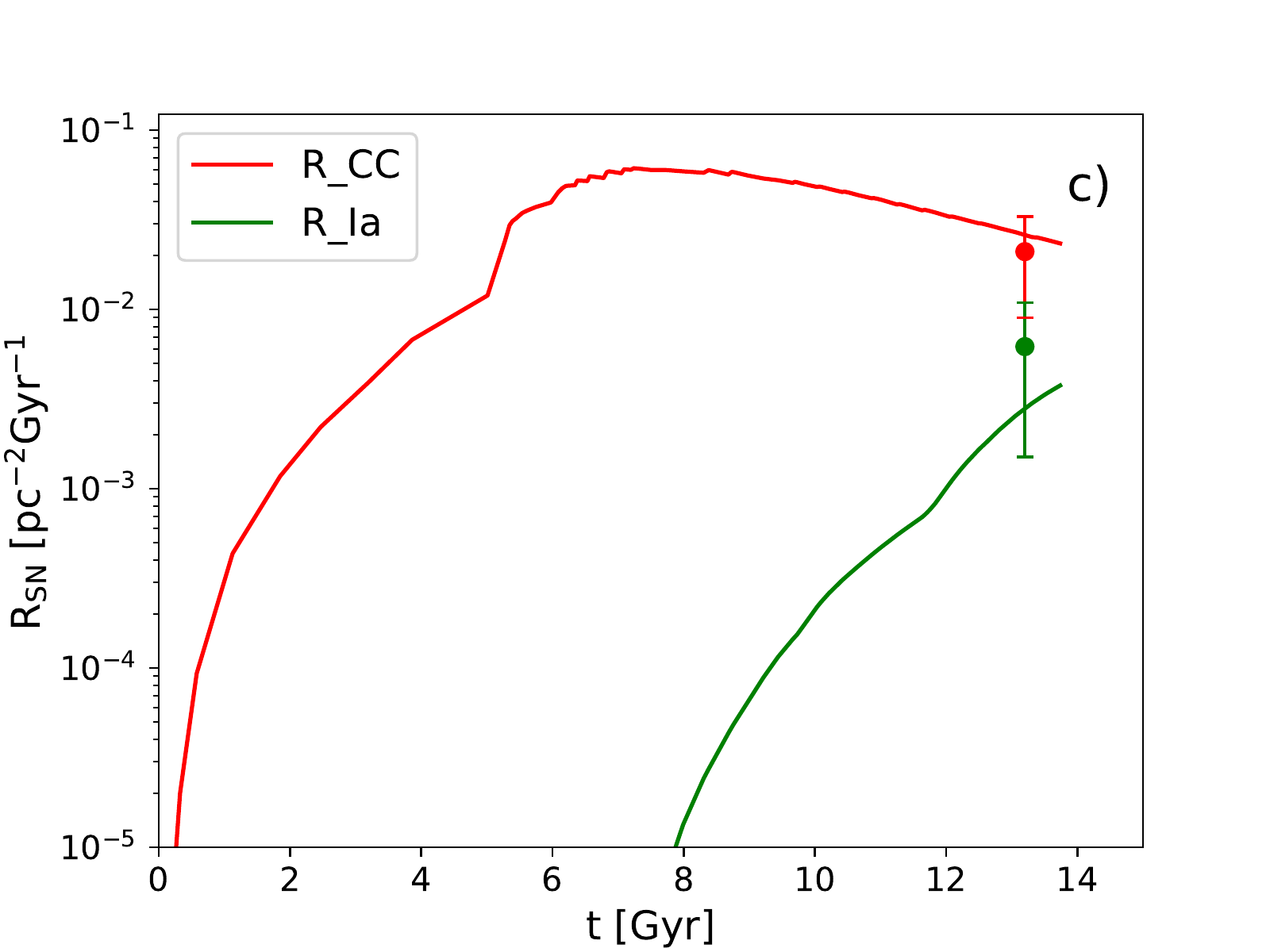}
\includegraphics[width=0.49\textwidth,angle=0]{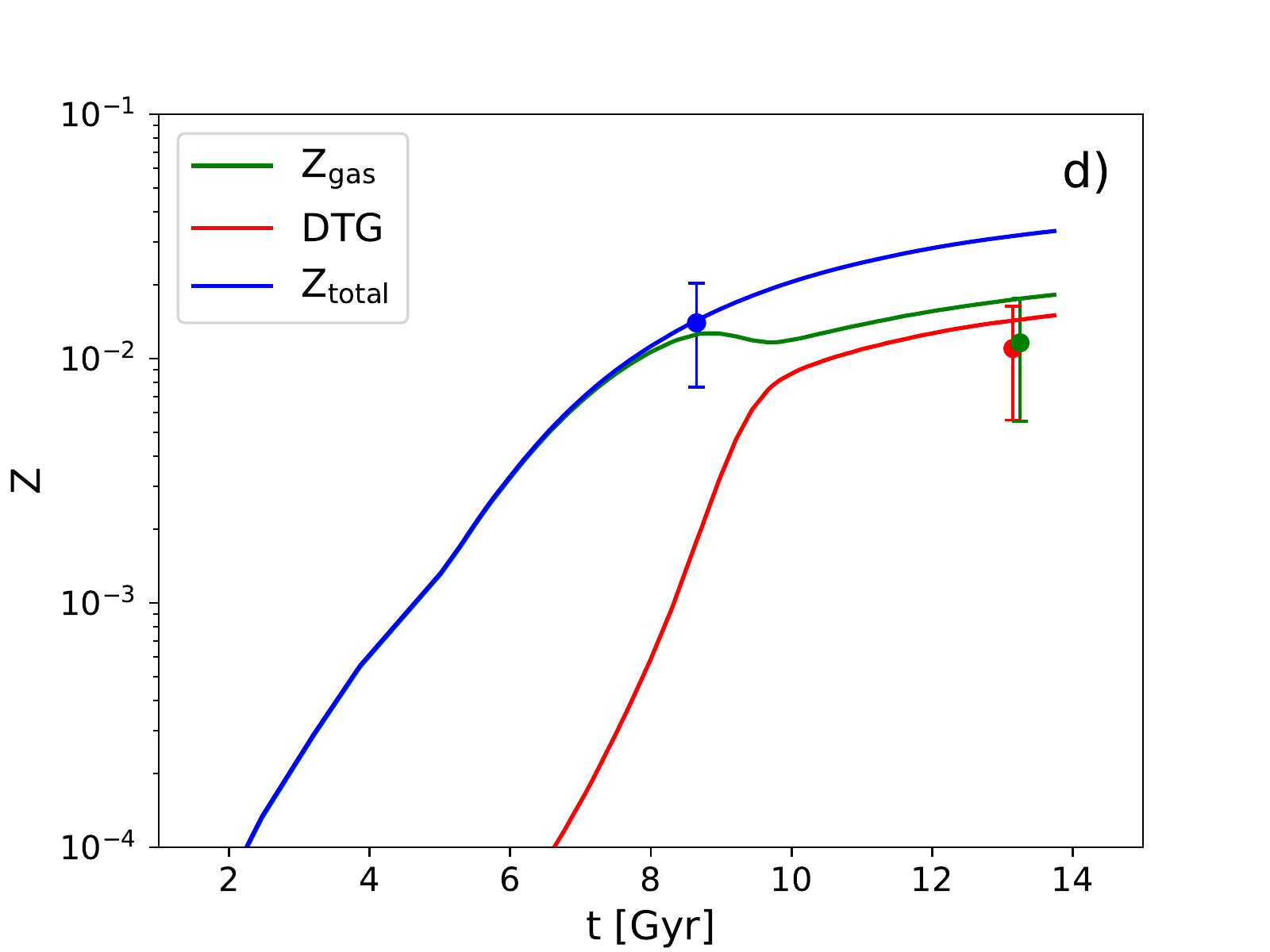}
\includegraphics[width=0.49\textwidth,angle=0]{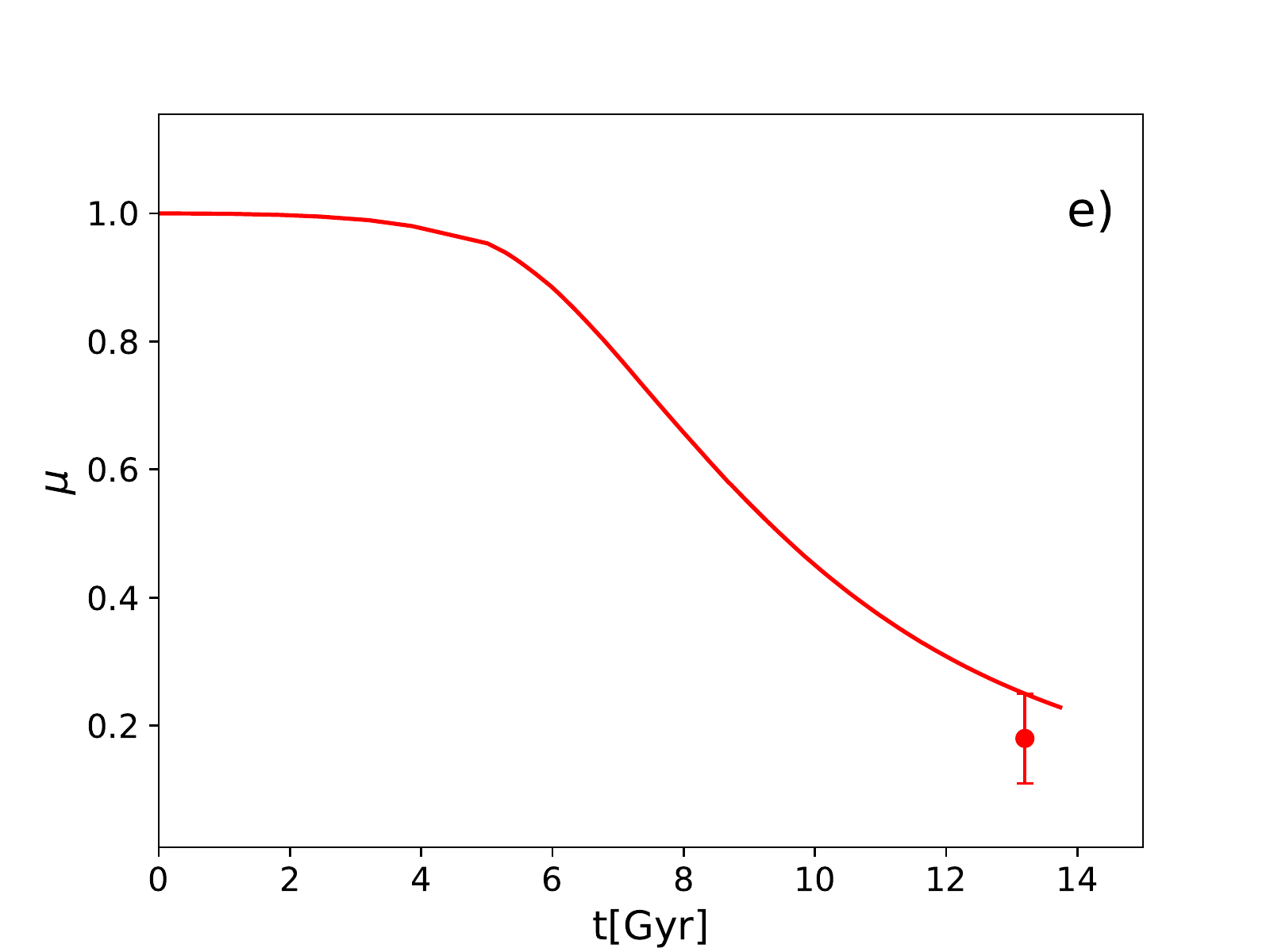}
\includegraphics[width=0.49\textwidth,angle=0]{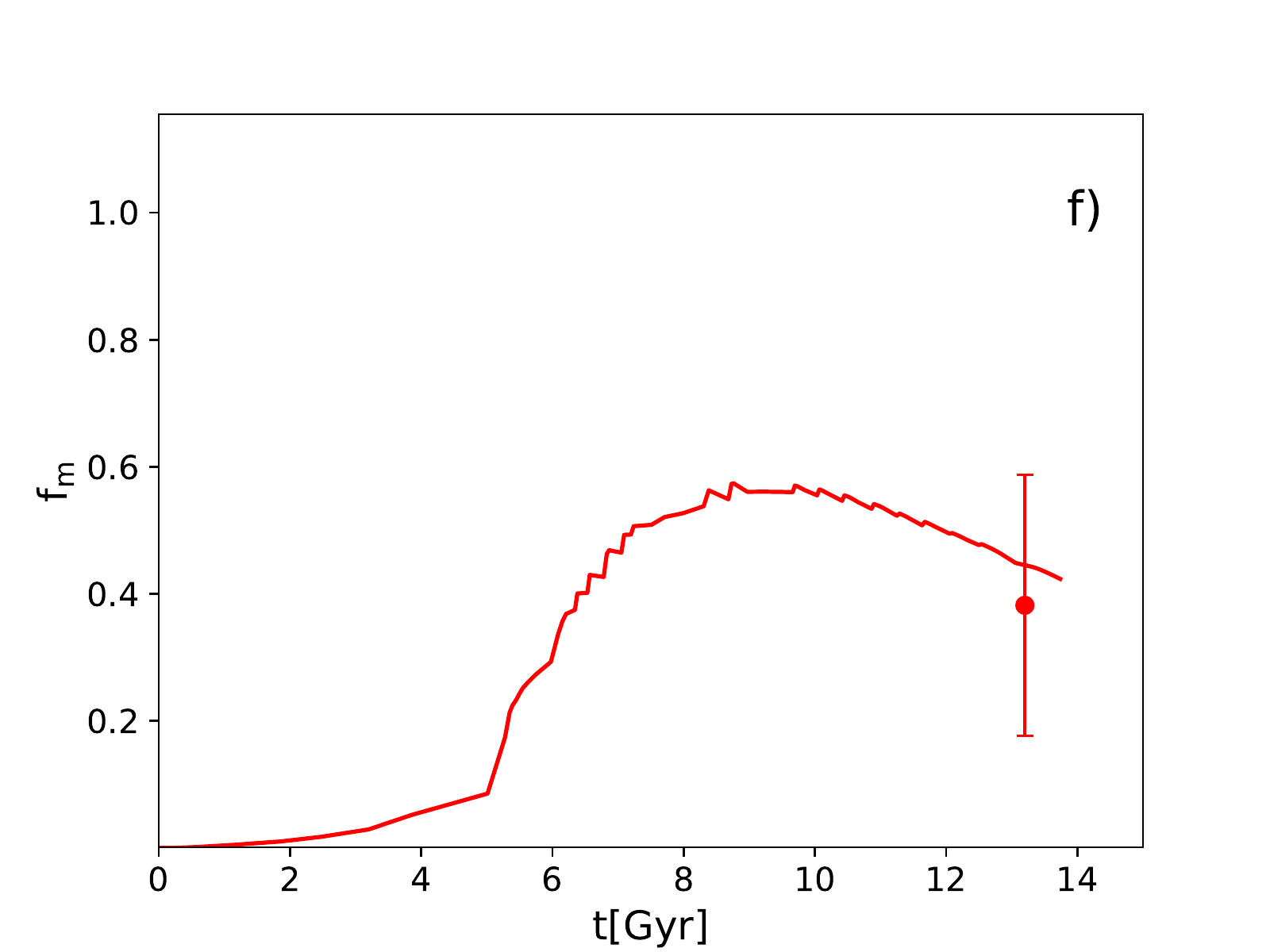}
\caption{Time evolution of the MWG-type galaxy model: a) Total mass density, $\Sigma_{\mathrm{total}}$, and stellar mass density, $\Sigma_{*}$; b) SFR; c) supernova rates; d) metallicity of the gas, $\rm Z_{\rm gas}$, dust to gas ratio, $\rm Z_{\rm dust} = DTG$, and total metal content of the ISM, $\rm Z_{\rm tot} = Z_{\rm gas} + Z_{\rm dust}$; e) Gas fraction, $\mu$, and f) molecular fraction of the ISM, $\rm f_{\rm m}$. The observational data are resumed in Table \ref{tab_obs}.}
\label{Fig:time_evolution_Solar_Neighbourhood}
\end{figure*}

We analyse the time evolution of the key characteristics of our reference MWG model,in
Fig.~\ref{Fig:time_evolution_Solar_Neighbourhood}.  Panel a)
represents the total and stellar mass densities. This depends on the
assumed infall rate and final $\rm \Sigma_{\mathrm total}$.
\begin{figure}
\centering
\includegraphics[width=0.43\textwidth,angle=0]{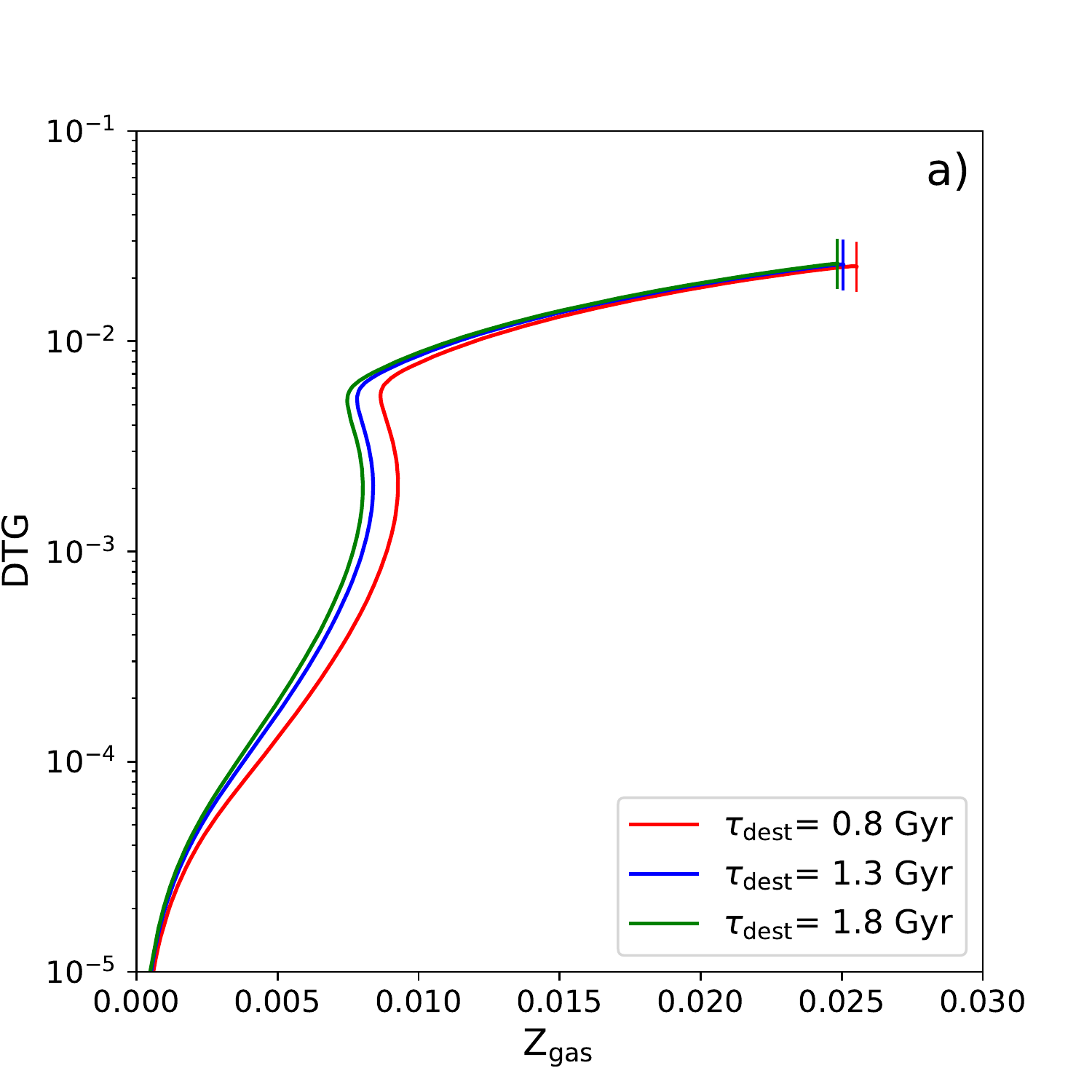}
\includegraphics[width=0.43\textwidth,angle=0]{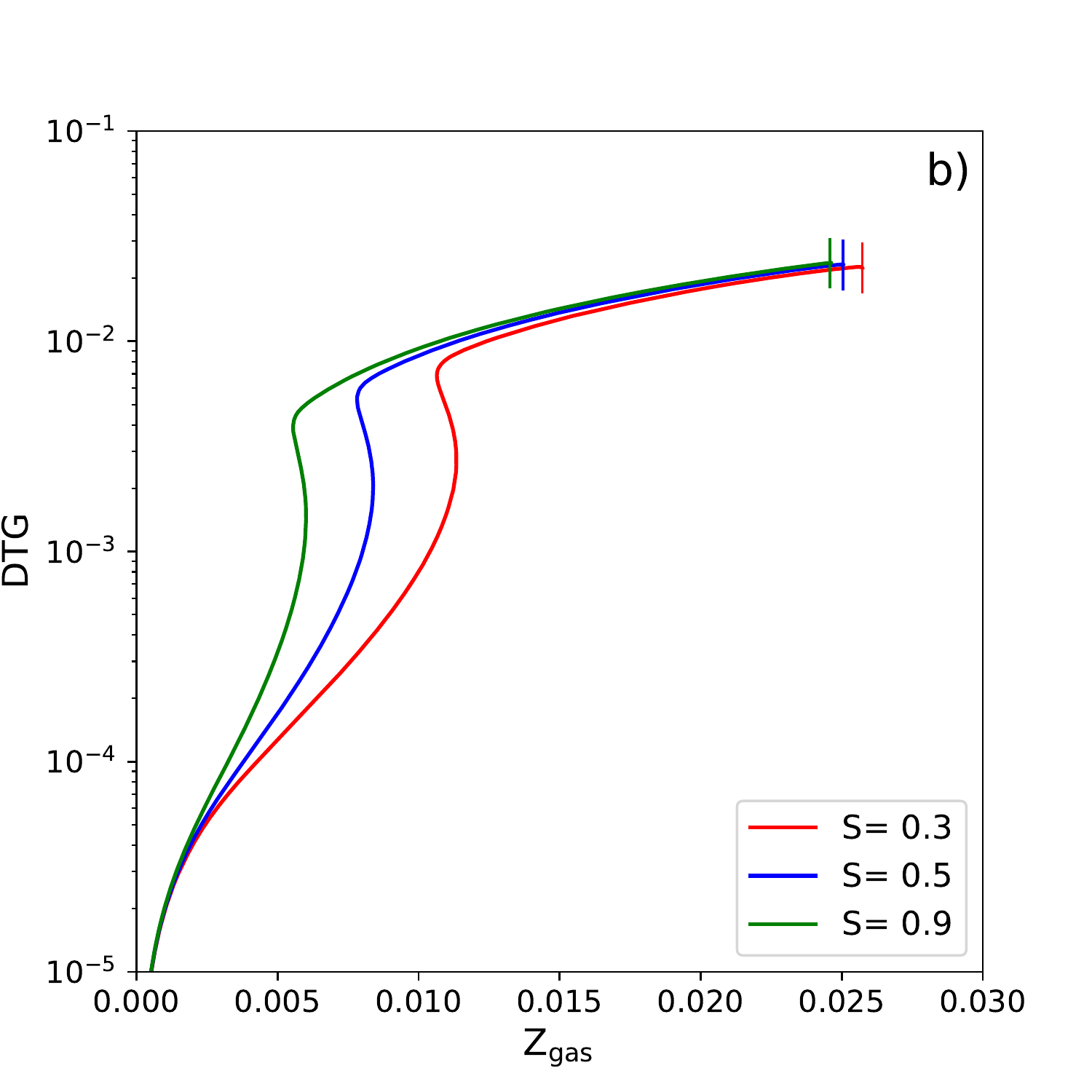}
\includegraphics[width=0.43\textwidth,angle=0]{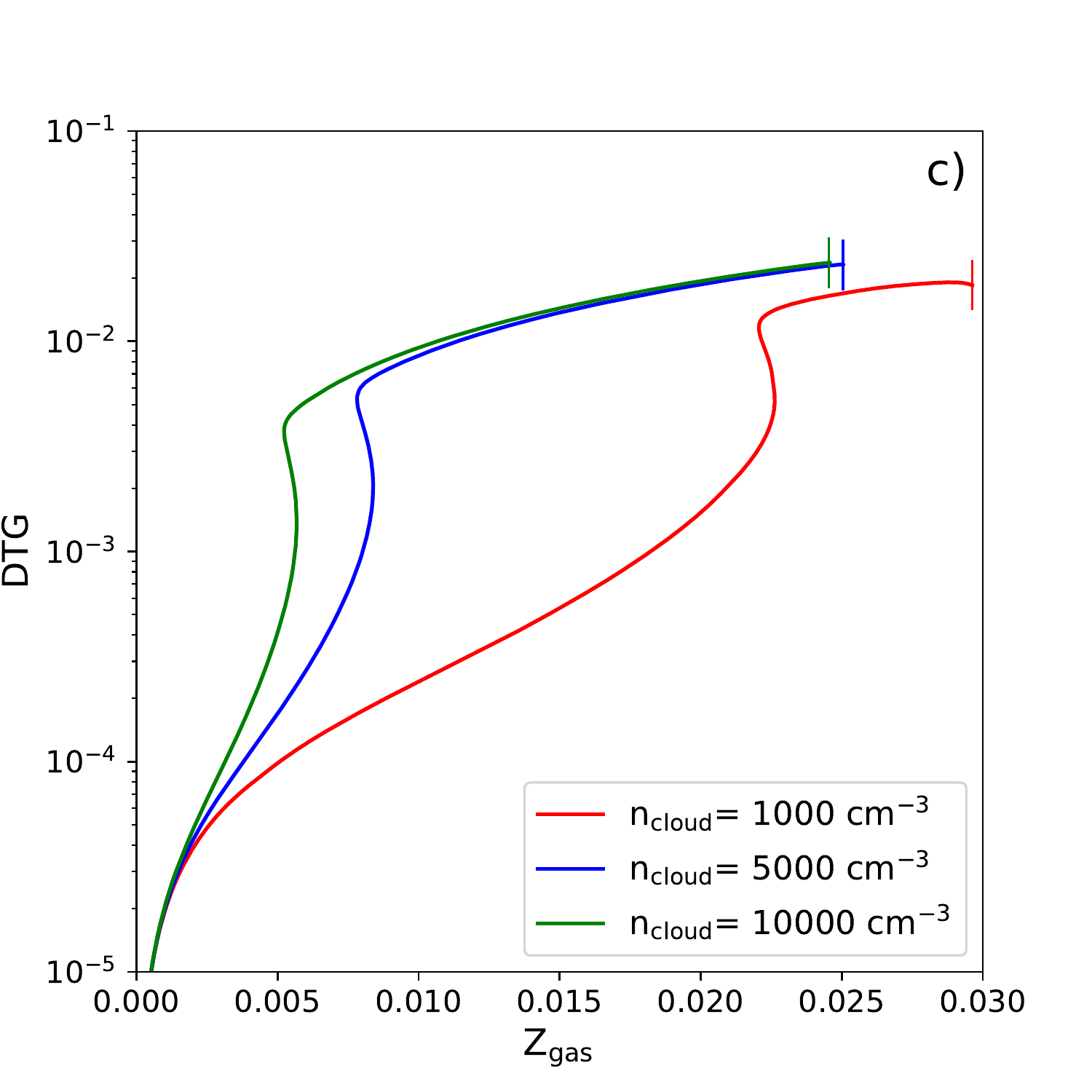}
\caption{Effect of the free parameters on the relation between DTG and $Z_{\rm gas}$: 
a) dust destruction timescale $\tau_{\rm dest}$; 
b) Sticking coefficient $S$; 
and c) number density of the molecular clouds $n_{\rm cloud}$.}
\label{Fig:DTG-Z}
\end{figure}

\begin{figure}
\centering
\includegraphics[width=0.43\textwidth,angle=0]{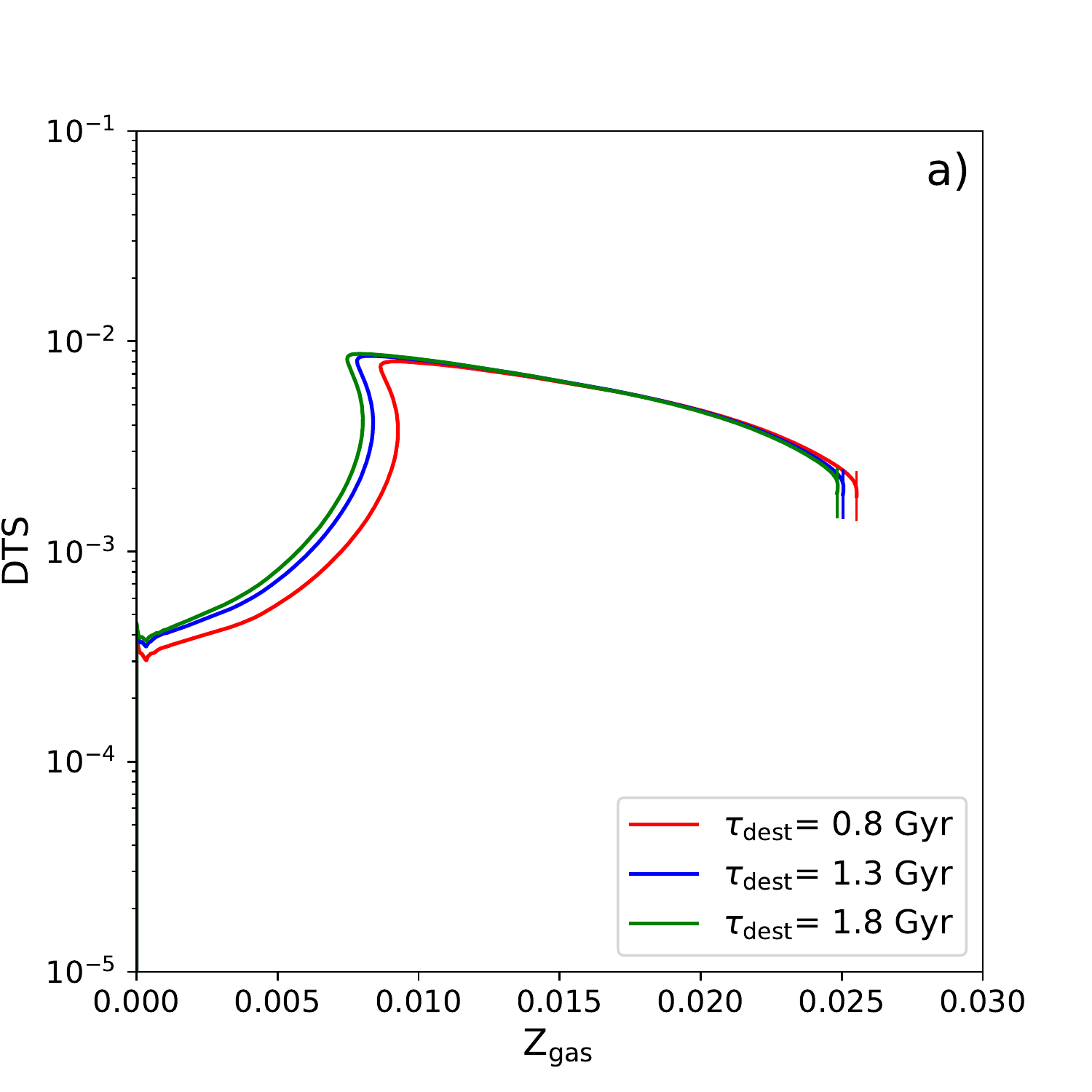}
\includegraphics[width=0.43\textwidth,angle=0]{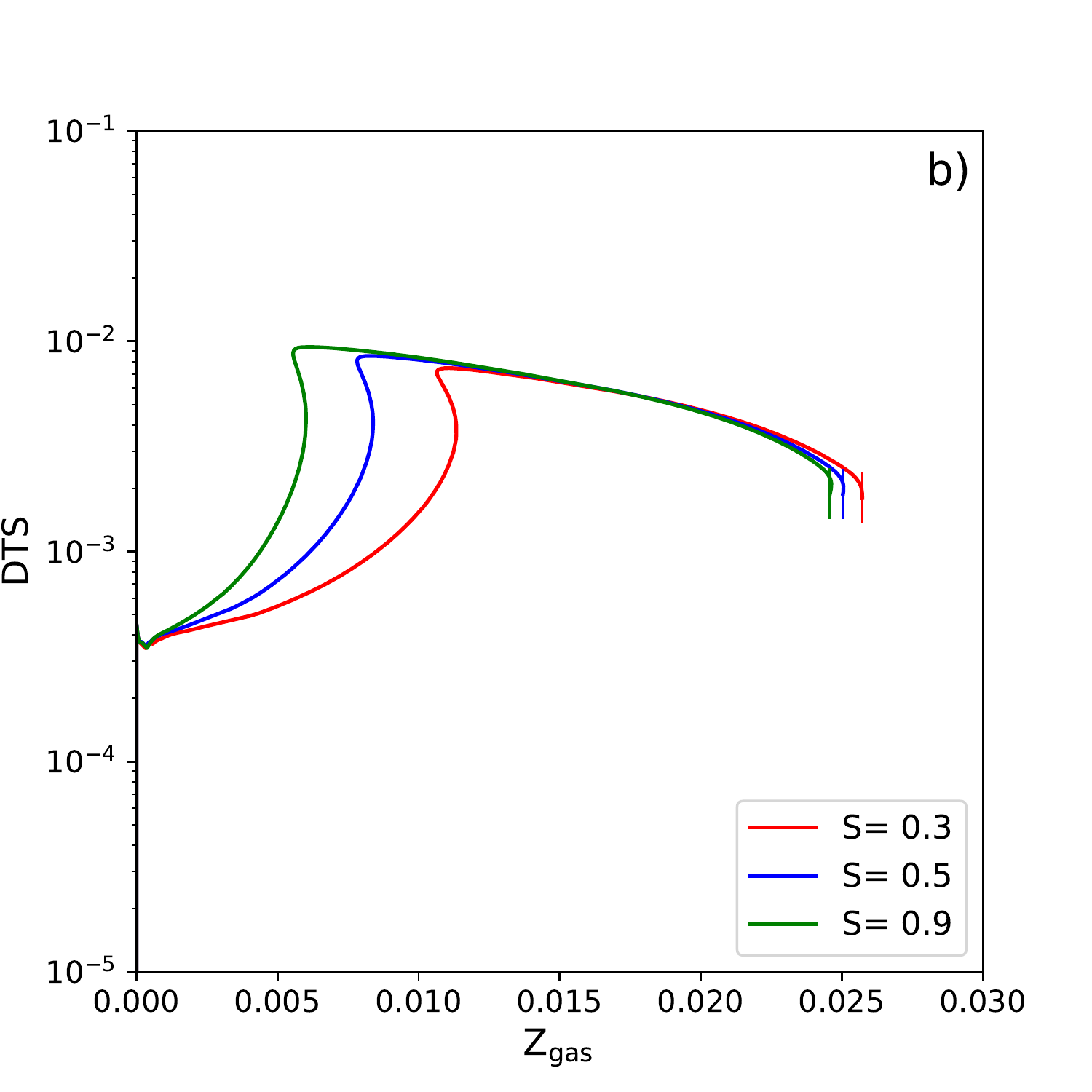} 
\includegraphics[width=0.43\textwidth,angle=0]{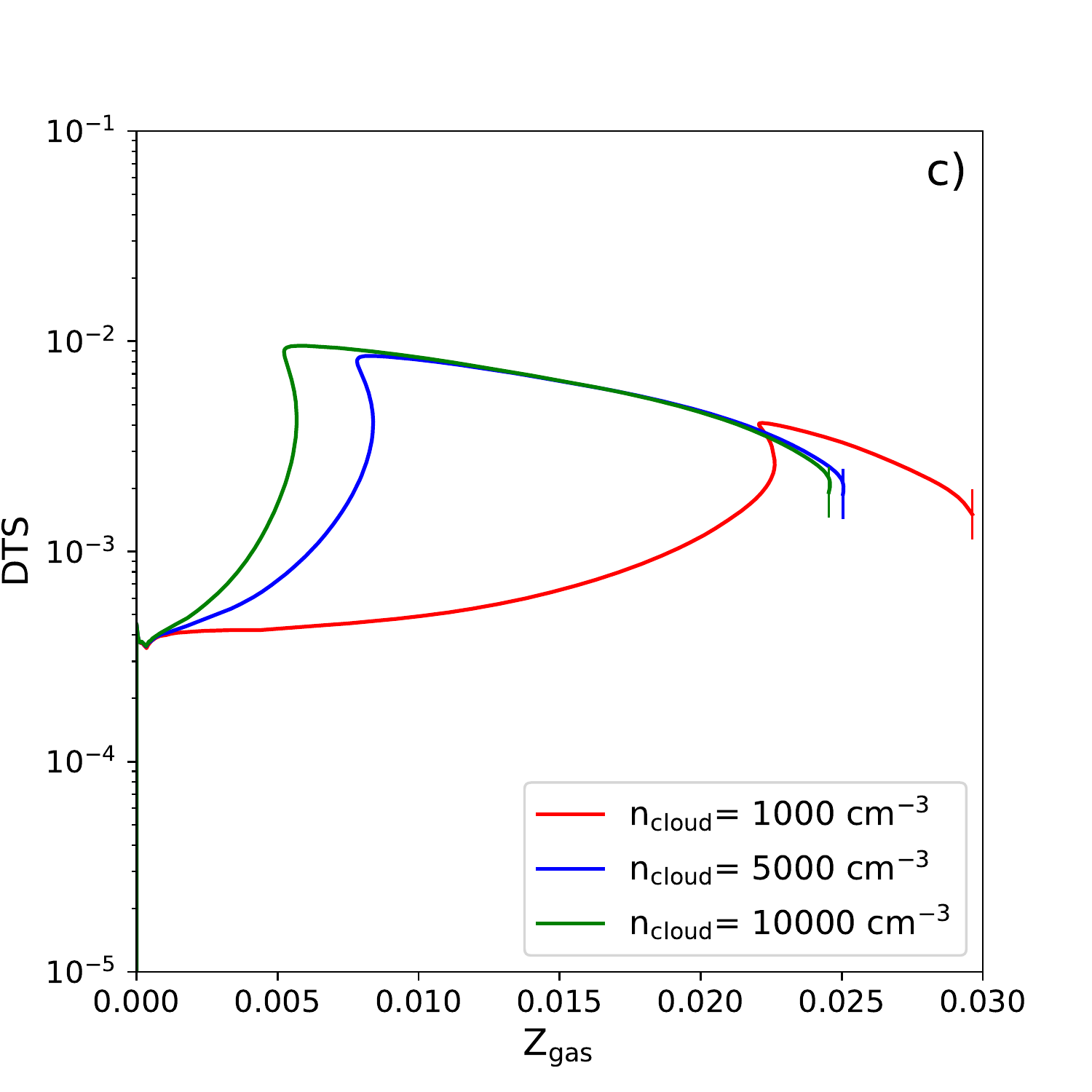}
\caption{Same as Fig.~\ref{Fig:DTG-Z}, for the dust to stars fraction.}
\label{Fig:DTS-Z}
\end{figure}

The SFR, shown in panel b), has a maximum, which appears at $\rm t\sim 5$\,Gyr,
decreasing smoothly afterwards until the present time, where a value
of $\sim 5\,M_{\sun}\,\rm pc^{-2}\,Gyr^{-1}$ is reached. This maximum
appears later than expected relative to the Milky Way's solar neighbourhood, due to our delay to form
stars as the new models 
require dust and molecular clouds to be created first, before stars can 
form.  On other hand, it is also necessary to remember that
our model is for the whole galaxy, thus it includes both the inner and outer regions of the disk, and, as such, represents an averaged behaviour, while  data come  from the
solar vicinity or regions located between 5-10\,kpc.

In panel c) we have plotted the SNe rates, both core collapse and type
Ia. Core collapse supernovae have a maximum at $ \rm t \sim 5 \ Gyr$, after which it decreases until it reaches $\rm R_{CC} \sim 2
\cdot 10^{-2} \ pc^{-2} Gyr^{-1}$. Type Ia SNe also reach the observed
values within the error bars, $\rm R_{Ia} \sim 2 \cdot 10^{-3}
\ pc^{-2} Gyr^{-1}$.

In panel d), we have plotted the time evolution of the metal
enrichment of the galaxy, keeping track of the metallicity of the gas
as well as the metals depleted in dust grains. The metallicity
grows with time until it reaches a critical value of $\rm Z \sim 0.3
Z_{\odot}$. At this critical metallicity, the accretion of dust begins
to dominate the evolution in the ISM and the metallicity of the gas
decreases somewhat as the dust grains accrete metals faster than they
are ejected by stars. However, it can be observed that the total
metallicity (including metals in gas and dust phase) continues
increase throughout the evolution of the galaxy. Both phases display
a similar evolution after $\sim 8$\,Gyr with an approximately constant
depletion factor of $0.3-0.4$, as observed. Thus, the total metallicity
is about twice the one observed in the gas phase, reaching a
supersolar value of 0.035 ($\sim 2.5 Z_{\sun}$).

Panel e) shows the fraction of gas, $\mu$ which must finish with a
value around a $15\%$; and in panel f), the molecular gas fraction
$\rm f_{m}$ is shown to be negligible at the beginning of the galactic
evolution, until the production of molecular hydrogen without dust
grains creates a critical amount, which pushes on the evolution of the
galaxy. The molecular fraction reaches a maximum value of $0.7$ at
$\rm t \sim 6$\,Gyr (a little after the maximum seen in the SFR) and from that
point it decreases gradually, as the reservoir of molecular gas is
being consumed to create stars. At the end of the evolution, the
molecular fraction reaches $\rm f_{m}\sim 0.45$, within the observed
range.

\subsection{Variation of the dust parameters}\label{subsec:Free_parameter_variation}

In order to test dependence of the results on the different free
parameters of the model, we now vary the adopted values for the
sticking coefficient (S), the destruction timescale
($\tau_{\mathrm{dest}}$) and the mean number density of the molecular
clouds ($\mathrm{n}_{\mathrm{cloud}}$).  We will see how changes in
each free parameter affect the relations between the the dust to gas
ratio (DTG) and dust to stars ratio (DTS) with the gas-phase
metallicity ($\rm Z_{gas}$) in the MWG model. The parameter range
covered in the total set of models is given in
Table~\ref{tab_run_models}. To see the effect of each free parameter
individually, we use a subsample of this grid, adopting three values
for each parameter.

\subsubsection{DTG-Z relation}\label{subsubsec:DTG_Z}

First, we illustrate the effect of modifying the destruction
timescale, $\tau_{\rm dest}$. As the destruction of the dust by
astration just depends on the SFR, it is similar in all the models
with the same $\Sigma_{\rm total}$ and $\tau$ (which regulate the
SFR). Thus, the destruction via SN shock waves, thermal sputtering and
radiative torques, modelled by the destruction timescale, is going to
play a role on the equilibrium abundance of the dust. As can be
observed in panel a) of Fig.~\ref{Fig:DTG-Z}, if the destruction
timescale is higher/lower, the turning point of the DTG-Z diagram
shifts to lower/higher metallicities, but the effect is very slight
for the adopted parameter range, as we will see, in comparison with
the effect of the two other parameters, sticking factor $S$ or density
$n_{\rm cloud}$.

The consequences of modifying the sticking coefficient $\rm S$ are
explored in panel b) of Fig.~\ref{Fig:DTG-Z}. As seen before, the
DTG-Z relationship has a very similar shape for almost all the
combinations of free parameters: a first part with a quick growth up
to a inflection point, where the relation becomes a relatively flat
power law. The variation of the sticking coefficient changes the
position of the inflection point to higher/lower metallicity as the
sticking coefficient decreases/increases. The inflection point is the
critical metallicity where the accretion begins to overcome both dust
destruction processes and stellar ejecta. When the accretion becomes
the dominant process of dust mass gain, the DTG increases roughly
linearly with respect to $\rm Z_{\rm gas}$.

Finally, the number density of molecular clouds ($\rm n_{\rm cloud}$)
is critically involved in the accretion process and it plays an
important role in the equilibrium between dust creation and
destruction. Its effect in the DTG-Z relation is the most noticeable
of the three free parameters that we have analysed, as can be observed
in panel c) from Fig.~ \ref{Fig:DTG-Z}. When we increase
$\mathrm{n}_{\rm cloud}$, the metallicity of the turning point of the
DTG-Z relationship decreases, since accretion dominates the evolution
of the dust surface density at earlier times in the history of
galaxy. The change now is stronger than the previous ones.

\subsubsection{DTS-Z relation}\label{subsubsec:DTS_Z}

Another interesting track to follow is the relation between the DTS
and $Z_{\rm gas}$. As was the case for DTG-Z, the DTS-Z relation can be divided in
three different regimes: a) at low metallicity, the DTS displays a
strong growth up to the metallicity where the DTS is maximum, $Z_{\rm
  max\,DTS}$; b) at intermediate metallicity, the DTS decreases slowly
with Z; c) finally, at high metallicity, the DTS decreases abruptly
(this last phase is not always apparent in our models).

We have analysed the effect of the destruction timescale in panel a)
of Fig.~\ref{Fig:DTS-Z}. As happened in the DTG-Z relationship, the
effect of $\tau_{\rm dest}$ is a second order effect, negligible
compared to the effect of other free parameters.  In panel b) of the
same Fig.~\ref{Fig:DTS-Z} we see how the change in the sticking
coefficient $S$ affects to the DTS-Z diagram:if $S$ increases, it
changes the value of $Z_{\rm max\,DTS}$ to lower values, but located
at higher metallicities. Again, as in the DTG case, this parameter
produces important variations in the relations of the dust with other
phases when is modified. Finally, we show the variations in the DTS-Z
diagram when we modify $\mathrm{n}_{\rm cloud}$ in panel c) of
Fig.~\ref{Fig:DTS-Z}. The increase of $\mathrm{n}_{\rm cloud}$
enhances the accretion rates of dust, decreasing the value of $Z_{\rm
  max\,DTS}$.

\begin{figure*}
\centering
\includegraphics[width=0.48\textwidth,angle=0]{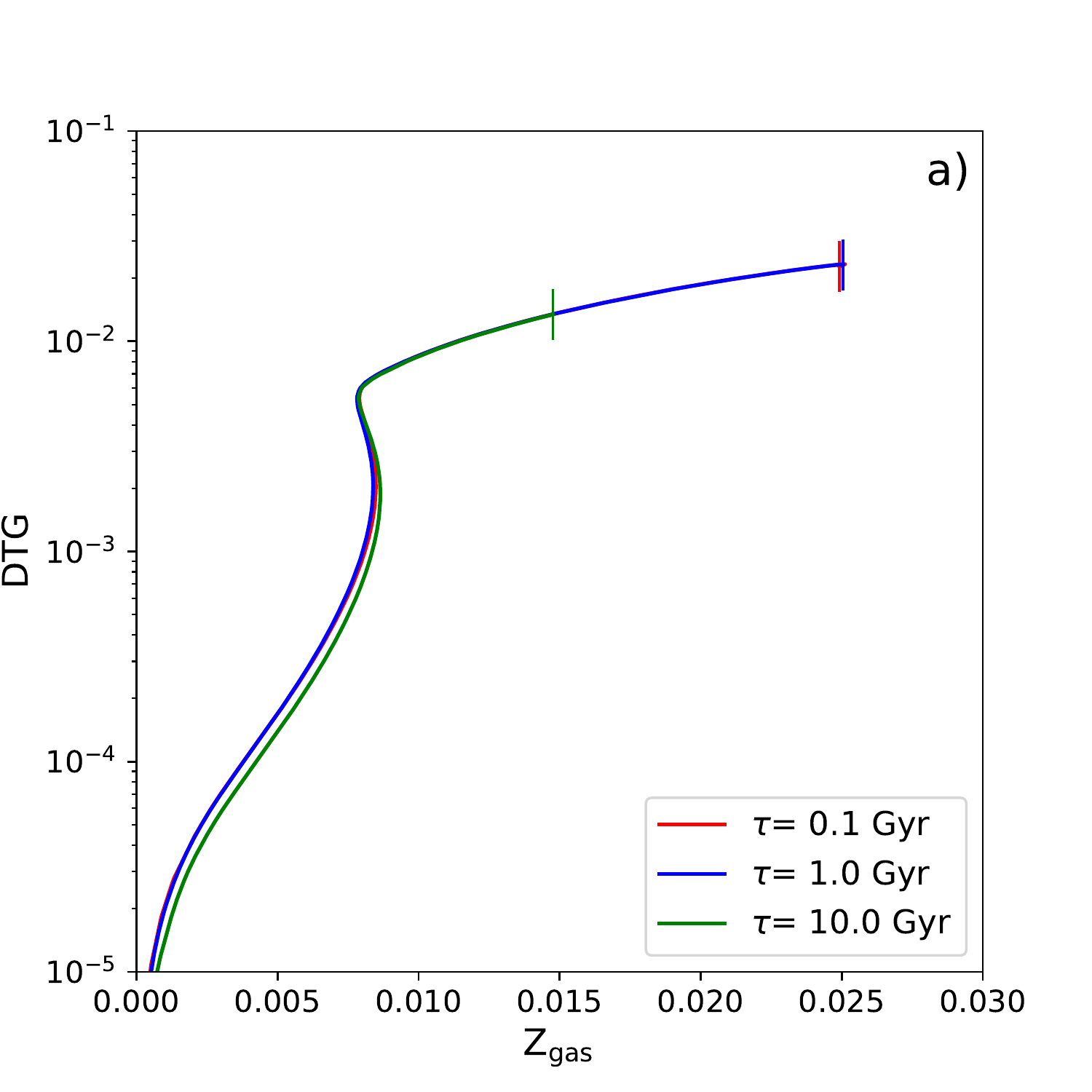}
\includegraphics[width=0.48\textwidth,angle=0]{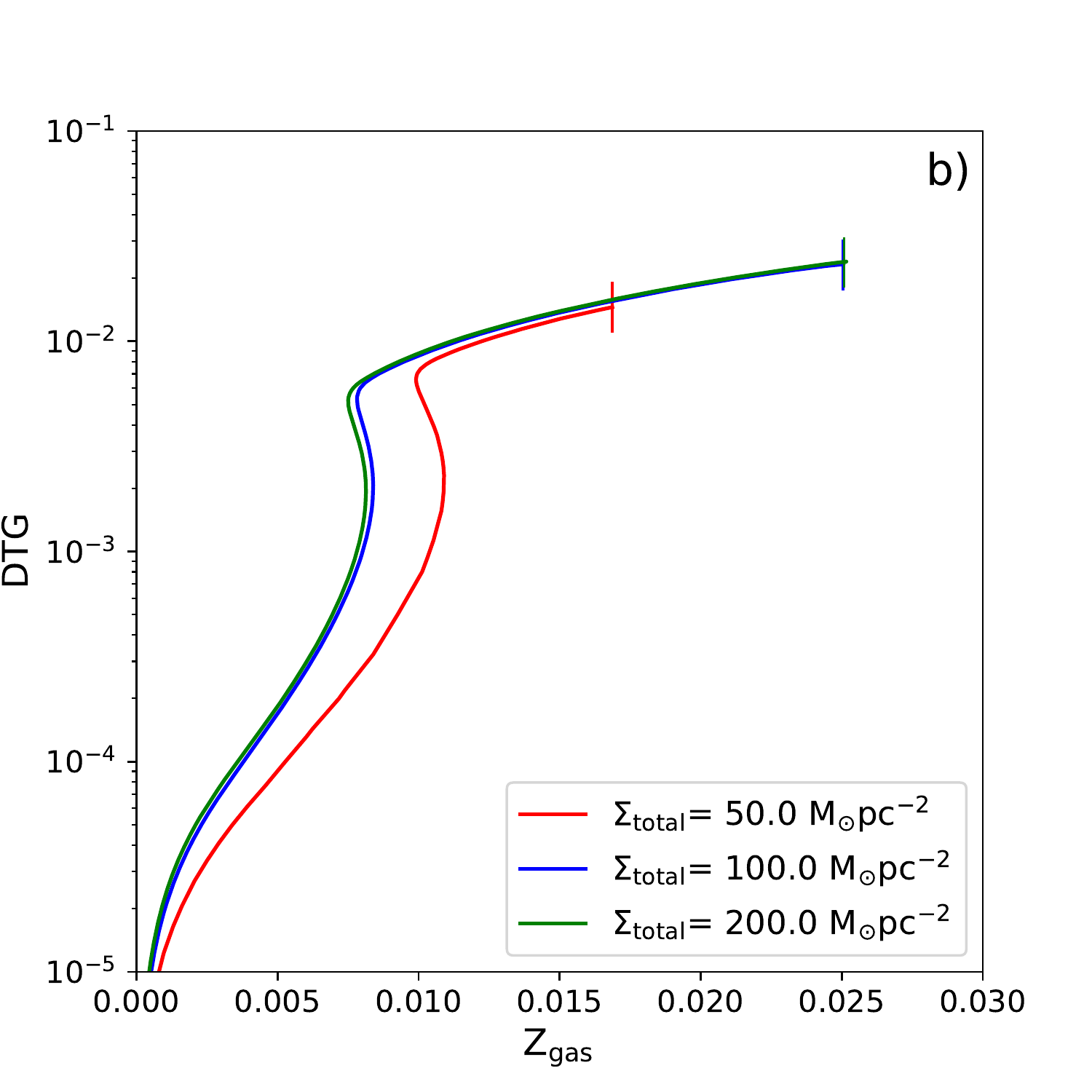}
\includegraphics[width=0.48\textwidth,angle=0]{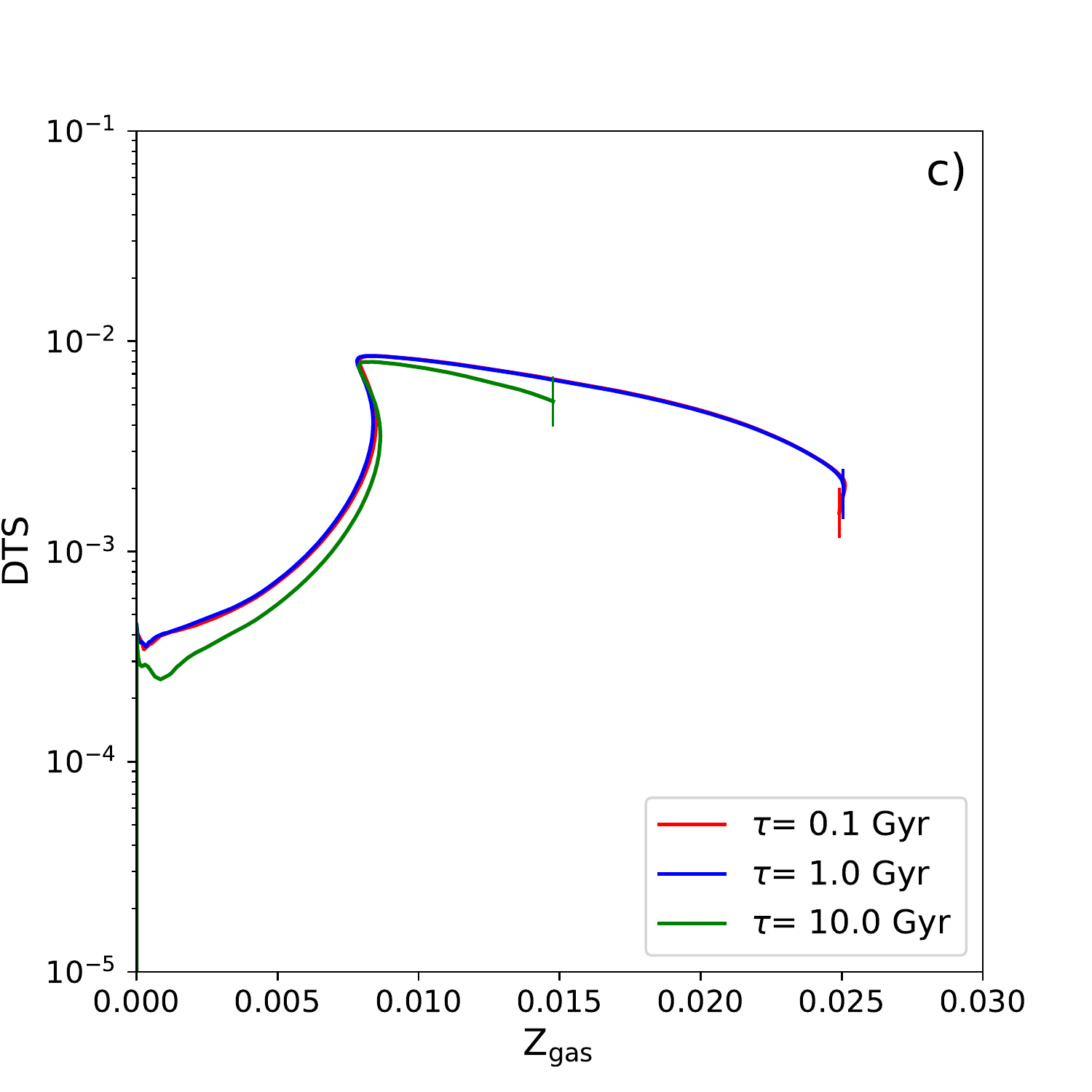}
\includegraphics[width=0.48\textwidth,angle=0]{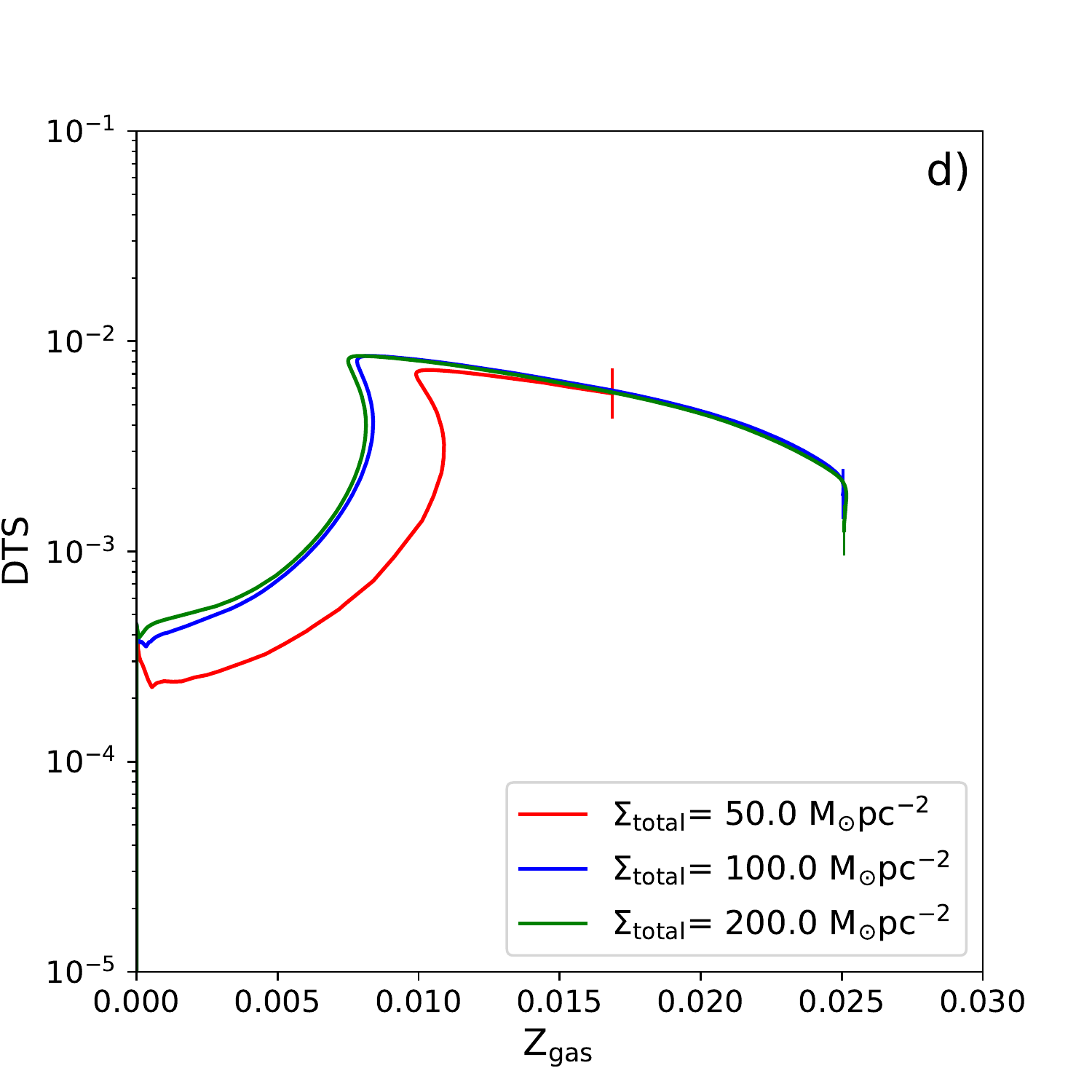}
\caption{ Effect of the different infall rates on (top panels) the
  relation DTG gas-phase metallicity $Z_{\rm gas}$: a) infall timescale $\tau$; b) final total surface density $\Sigma_{\rm total}$; and (bottom panels) on the relation DTS gas-phase metallicity: c) infall  timescale $\tau$; and d) final total surface density $\Sigma_{\rm total}$.}
\label{Fig:DTG_DTS}
\end{figure*}
These latter curves are very much different from those seen in the previous panels; specifically, this particular parameter is very important in shaping the evolution of the region, although if $\mathrm{n}_{\rm cloud} \ge 5000 \,\rm cm^{-3}$, the final DTS will be very similar in all cases.

\subsection{Variation of the infall parameters}\label{subsec:Comparison}

In order to compare our models with observations of different galaxies, we must used other plausible surface densities and infall timescales.
We illustrate the effect of varying the infall timescale $\tau$ in the left panels a) and c) of Fig.~\ref{Fig:DTG_DTS}. As this parameter regulates the rate at which gas falls into the galaxy, it affects the time evolution of the dust, metals, gas and stars. Increasing/decreasing the infall parameter delays/enhances the evolution of the galaxy, reaching lower/higher values of the metallicity, DTG and DTS at any given time. However, both DTG-Z and DTS-Z relations remain unchanged, because the infall variations affects equally to the three variables (DTG, DTS and Z) involved in these figures. So, the only thing that changes $\tau$ is the speed at which the galaxy evolves, without changing the shape of the diagram, delaying the whole evolution of the galaxy when it is lengthy and enhancing it when it is shorter.

The effect of the final total surface density is seen in panels b) and d), with a clear difference between the model with $\Sigma_{\rm  total}=50\,\rm M_{\sun}\,pc^{-2}$ and the other two models. 
This result hints the existence of a saturation effect in $\Sigma_{\rm total}$, having similar results for DTG-Z and DTS-Z above a surface density limit of  $\sim \Sigma_{\rm total}=100 \rm M_{\odot}$. The latter implies that the transition from ejecta-dominated to accretion dominated regimes occurs at lower metallicities in high surface density regions than in low surface density ones. 

\subsection{Comparison with observations for spiral galaxies}\label{subsec:Data}
\begin{figure}
\includegraphics[width=0.49\textwidth,angle=0]{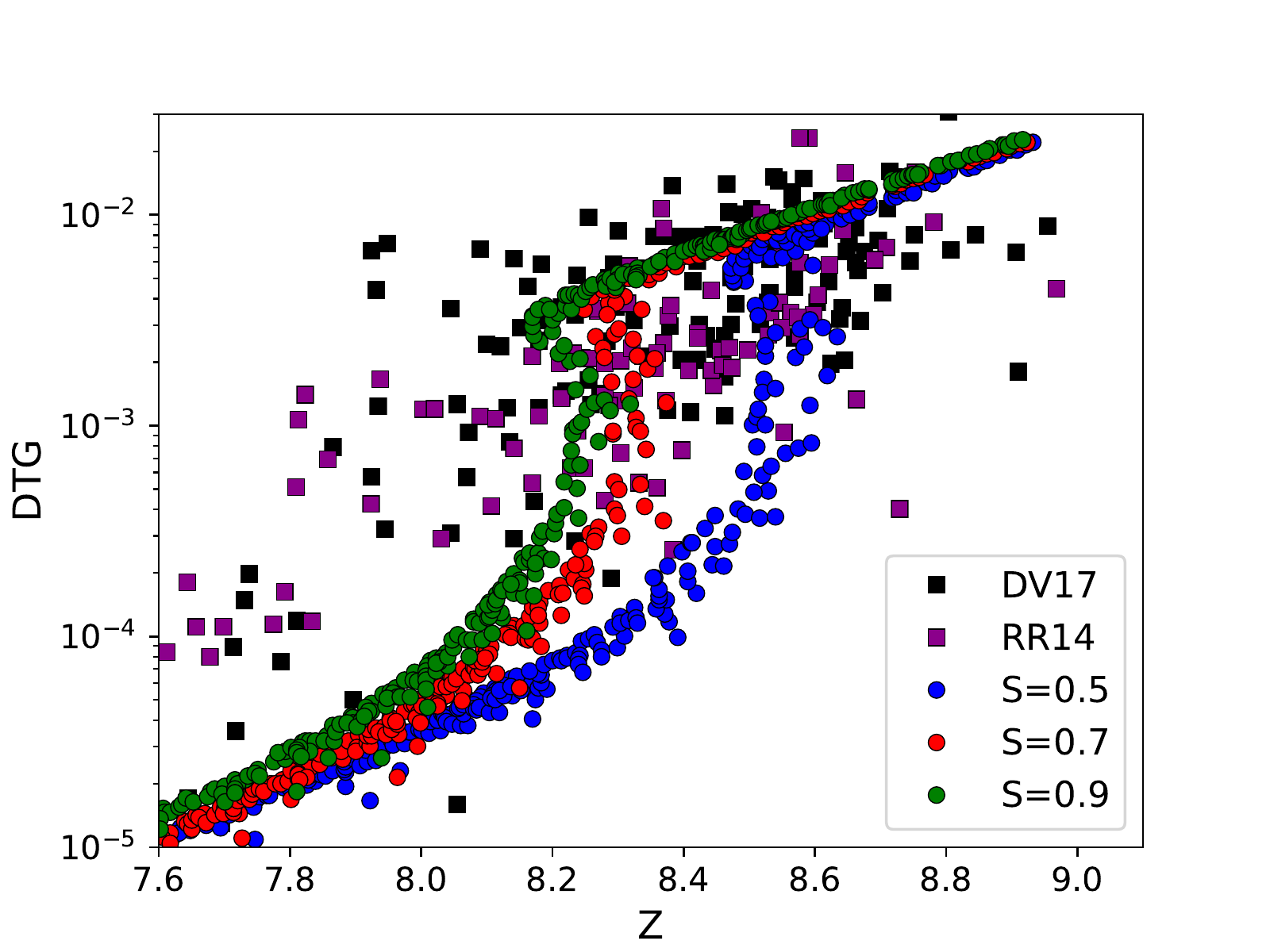}
\caption{Comparison of results of our set of models with the galactic surveys from RR14 and DV17.}
\label{Fig:Comparison_RR14_and_DV17}
\end{figure}
We compare the obtained results with the most relevant studies of the DTG-Z relationship of the literature, namely, the surveys of galaxies of RR14 and \citet[][hereinafter DV17]{De_Vis2017}
\footnote{In order to compare with the observational data, we have converted our values of $Z_{\rm gas}$ to 12+log(O/H) using the solar values of AS09: $\rm Z=Z_{sun}\times 10^{12+log(O/H)-8.69}$.}. In order to test our models correctly with the galaxies of the nearby universe observed in these two surveys, we represent only the last time of each model. These models, which we have selected from our set of models, have a same density of the clouds, $\rm n_{cloud} = 5000 \ cm^{-3}$ and a same destruction timescale $\tau_{\rm dest} = 1.8\,\rm Gyr$.  We have chosen among these subset models, those with 3 different sticking coefficient, i.e., $\rm S =0.5$, $=0.7$ and $= 0.9$, combined with all range of infall inputs, $\tau$ and $\Sigma_{\rm total}$.

This comparison is shown in Fig.~\ref{Fig:Comparison_RR14_and_DV17}. It can be observed that our models reproduce values in the range of solar and super-solar metallicity; admittedly, some parameters reproduce better these data than others. Models that use free parameters more favourable to create dust grains, i.e., $\rm S = 0.9$, $\tau_{\rm  dest}=1.8\,\mathrm{Gyr}$ and $\mathrm{n}_{\rm cloud}=5000\,\mathrm{cm}^{-3}$, have a critical metallicity at $\rm 12+log(O/H) \approx 8.20$, similar to the value $\rm 12 + \mathrm{log}(O/H) \approx 8.10$ obtained by RR14. Below this limit, our models generally give DTG values below the observed ones.

We must consider that DV17 and RR14 do not take into account all components of the galaxies studied; e.g., DV17 does not consider the molecular gas. Conversely, RR14 does include this component, but they are not able to estimate the associated uncertainty attached to this important component. Such considerations means a degree of precision lost: if the molecular gas component would be included in the DV17 data, the DTG values would be lower, which may be specially important in the low metallicity region. The uncertainties in the data used to compute DTG may be high, since e.g. H{\sc i} is usually computed until a larger radius than the stellar profile or the oxygen abundances. Thus, the estimates of DTG for each galaxy as a whole are lacking in precision when compared with the predictions of our models.

\begin{figure}
\includegraphics[width=0.46\textwidth,angle=0]{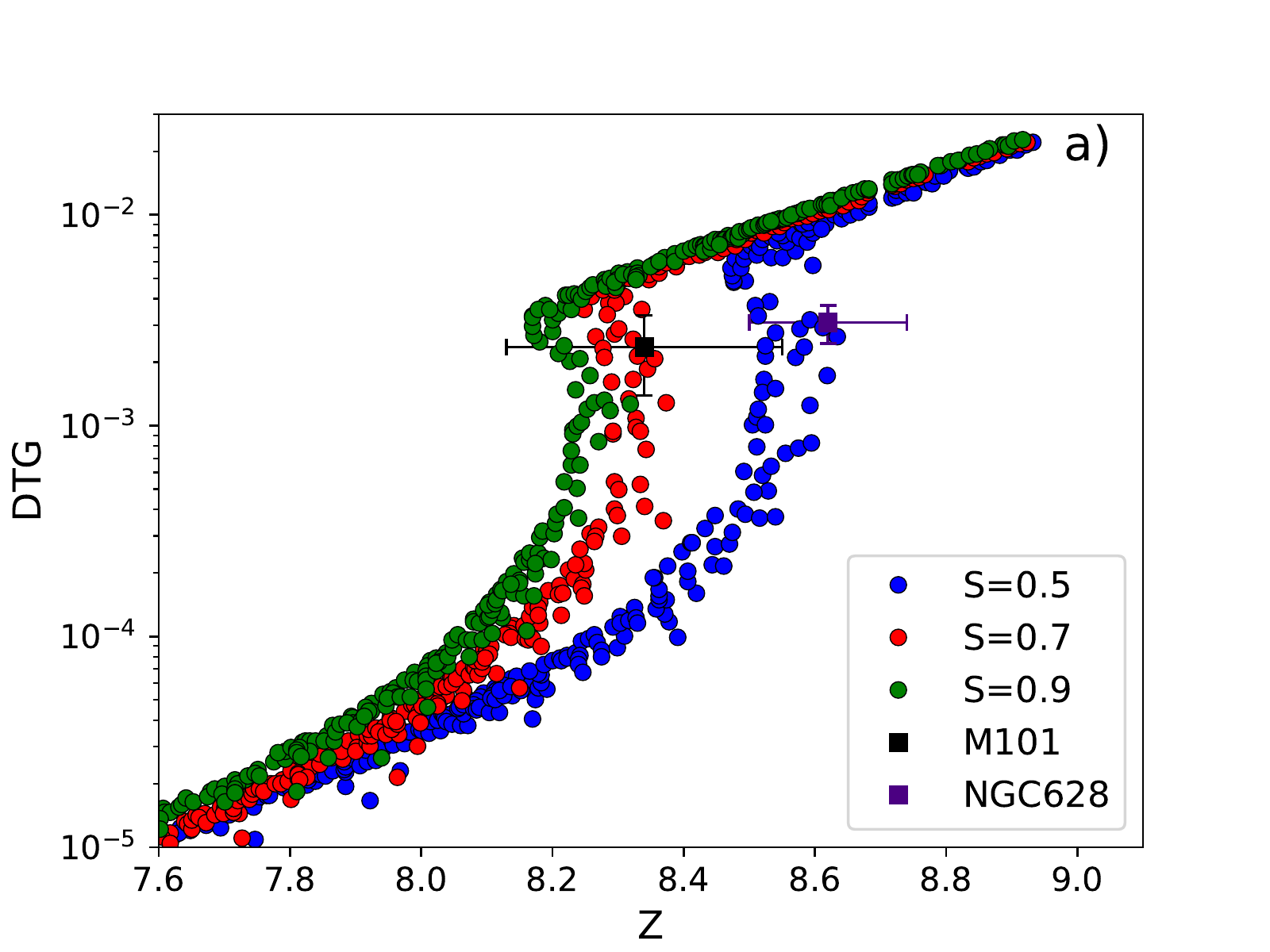}
\includegraphics[width=0.46\textwidth,angle=0]{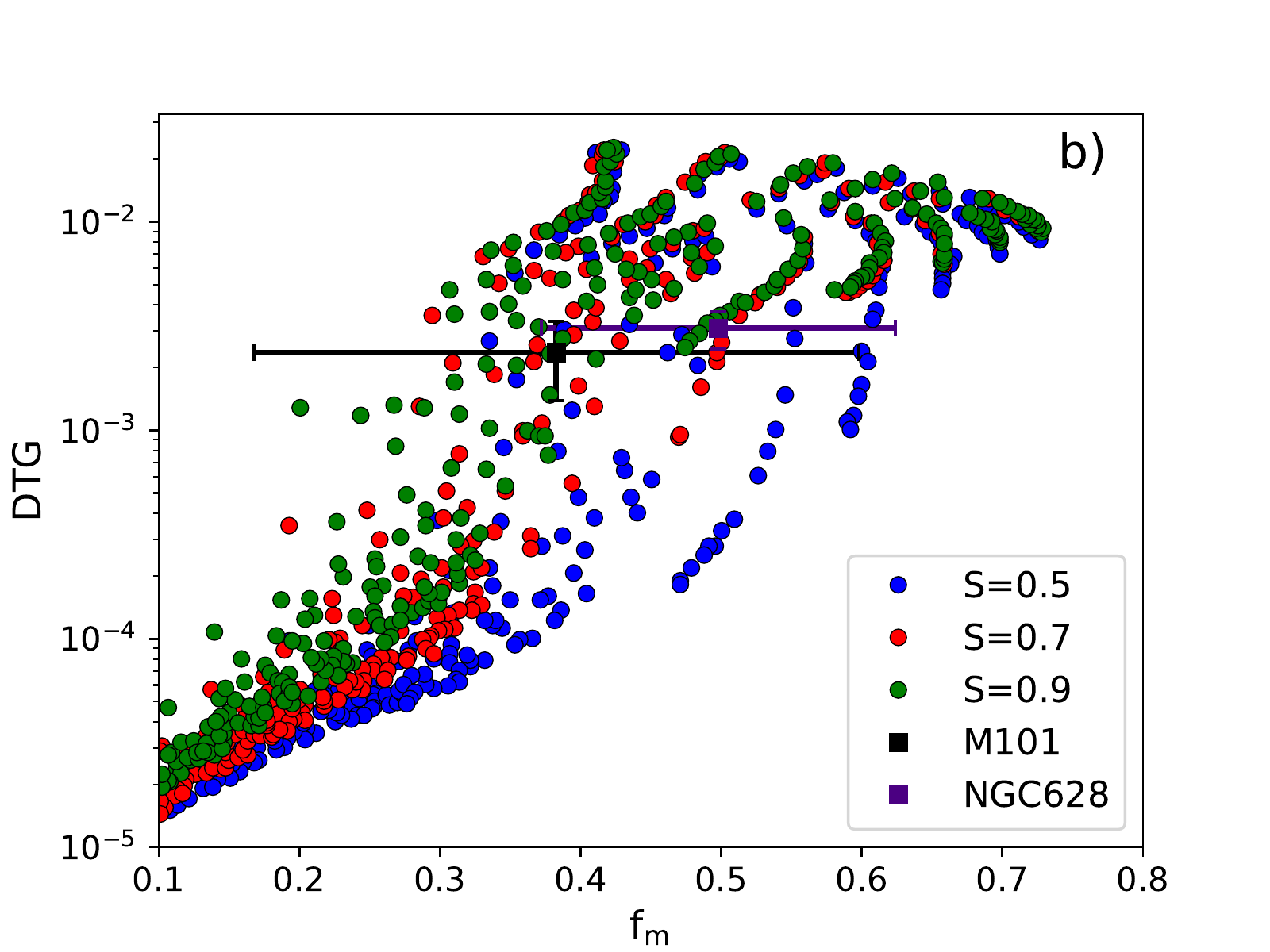}
\includegraphics[width=0.46\textwidth,angle=0]{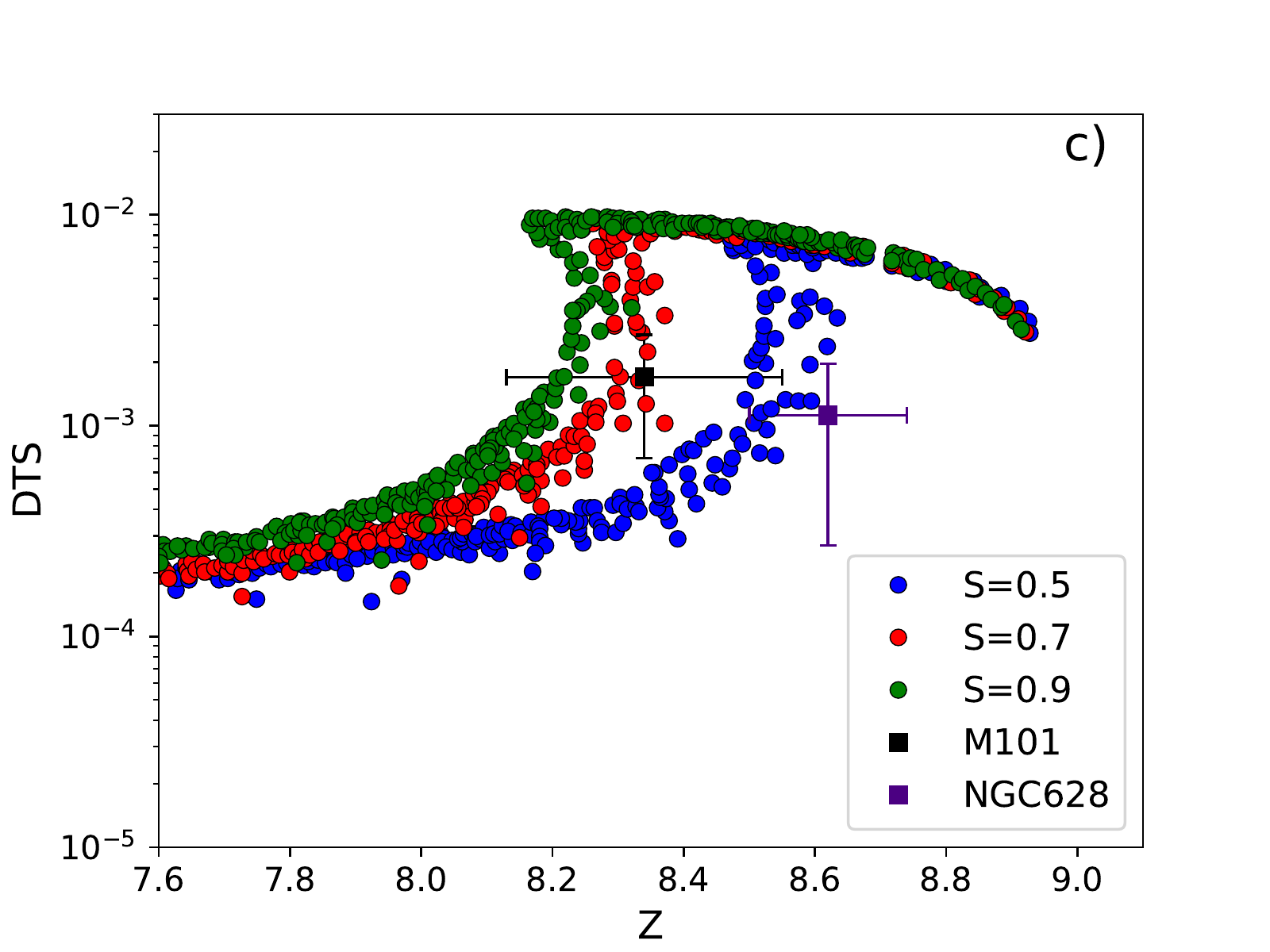}
\caption{Comparison of our models with the results of the observations for M~101 and NGC~628 from \citep{Vilchez+2019}: a) DTG {\sl vs} Oxygen abundance; b) DTG {\sl vs} the molecular gas fraction $\rm f_{m}$; and c) DTS {\sl vs} Oxygen abundance.}
\label{Fig:DTG_comparison_NGC628_M101}
\end{figure}

We compare now our results with the full data for two particular galaxies, M~101 and NGC~628, from \citet[][hereinafter V19]{Vilchez+2019}. We show the DTG-Z, DTS-Z and DTG-$\mathrm{f}_{\rm m}$ relationships in Fig.~\ref{Fig:DTG_comparison_NGC628_M101}, panels a), b) and c), respectively.
Since we have not computed different radial regions within each galaxy, and our models correspond to global data for whole galaxies, we have computed the mass of each gas phase and dust for each radial regions and we have computed the mean mass and standard deviation of HI, $H_2$, 12+log(O/H) and dust. 
This implies that there is a big dispersion of the mean values for the two galaxies. Both galaxies' data with their error bars lie over our models, suggesting that they have yet to reach the saturation level, and therefore more evolution remains possible. M~101 is consistent with a model possessing $\rm S=0.7$ while NGC~628 is best fit with $\rm S=0.5$. 

Overall, this comparison suggests that our models reproduce adequately the observations in medium-to-high metallicity regimes, but the low metallicity regime in the DTG-Z diagram is not well fit.

\section{Discussion and conclusions}\label{sec:Conclusions} 

Summarising, we have created a self-consistent model of galactic
chemical evolution that includes a multiphase ISM with three phases:
an ionised gas phase, which would correspond to a temperature of
$10.000\,\mathrm{K}$, a neutral gas phase, which would have a
temperature of $100\,\mathrm{K}$ and a molecular gas phase, whose
temperature would be $10\,\mathrm{K}$. This multiphase structure of
the ISM, is key to reproduce realistic star formation prescriptions via molecular
clouds, and we have obtained it without any free
parameters. 

Furthermore, we have included the life cycle of dust by
taking into account:
\begin{itemize}
\item The synthesis of dust grains in the ejecta of AGB stars, core-collapse supernovae and thermonuclear supernovae.
\item The reprocessing of dust in the ISM including the effect of destruction via SNe, thermal sputtering and radiative torques and the dust grain growth via accretion in the coldest phases of the ISM. 
\end{itemize}

In our formalism, it is only the dust processes, for which limited 
knowledge still exists, which need to employ free parameters. These 
parameters are related to the accretion phase within molecular clouds 
and with the destruction of dust: the sticking coefficient and the 
density of clouds for the accretion process, and the timescale of dust 
destruction.

With this model, we have computed a set of 1350 realisations for a
sample of different inputs of infall rate, combined with variations of
the three free parameters. Comparing our results with RR14 and DV17
data, two main conclusions come to light. On the one hand, the transition
between the low metallicity and the solar/super-solar metallicity regimes 
seems smoother in the observations than in our models. The latter
underestimate the DTG in the low metallicity range, $\rm 12 + \rm log
(O/H) \in [7.5,8.15]$. This discrepancy may be due to a more abrupt
than expected transition from ejecta-dominated to accretion-dominated
phases. Any underestimate in the error bars
associated with the observations will also contribute to this discrepancy. We must also consider the uncertainties remaining in the 
ejection tables employed, as the disparity between models and observations at low metallicity is dominated by said stellar ejecta.

Our conclusions can be summarised thusly:

\begin{enumerate}
\item The relationships of DTG-Z and DTS-Z while showing basically the  same shape, remain dependent upon the (quite unknown)
  sticking factor and cloud density,
\item The total stellar density has a saturation level when it reaches
  values higher than 100\,$\rm M_{\sun}\,\rm pc^{-2}$, after which the
  DTG-Z and DTS-Z remain invariant. 
\item Following our model, DTG increases very abruptly when a critical
  metallicity is reached. This critical Z depends on the free
  parameters, mostly the sticking parameter, but it does not disappear
  with any value of $\rm S$
\item Our models cannot fit the continuously decreasing DTG with
  decreasing Z shown at 12+log(O/H) < 8.15, since in this regime, our
  models show instead a steep variation. It is likely that other pathways to
  create molecular gas in the early phases of galaxy's evolution will be required,
  given the difficulty of forming stars when there is a low Z and a
  low dust ratio in our model. More work, including a new star
  formation prescription in the low Z regime appears necessary, although the relatively large
  uncertainties associated with the lower metallicity galaxies' data may also be
  a contribute to the disparity seen.
\item The models reproduce the observations located in the medium-to-high
  metallicity regime with variable sticking parameter. In particular,
  data from \citet{Vilchez+2019} for NGC~628 and M~101 are very
  reproduced for our models. These last data are provided as a function of galactocentric radius for these disks and, therefore, our next objective will be  to introduce 
  a spatially-resolved multi-zone
  approach into our modeling, each with the corresponding variable
  free parameters. This will allow a more self-consistent direct comparison
  with extant observational data.

\end{enumerate}

\section*{Acknowledgements}
This work has been partially supported by MINECO-FEDER-grants AYA2013-47742-C4-4-P, AYA2016-79724-C4-1-P and AYA2016-79724-C4-3-P. YA is supported by contract  RyC-2011-09461 of the \emph{Ram{\'o}n y Cajal} programme. 
The authors acknowledge the anonymous referee for her/his comments which have substantially improved the manuscript.
We thank to B.K. Gibson for a review of our manuscript.




\appendix 

\section{Particle density}
\label{sec_density}

One of the quantities that is widely used in our model is the particle density of each individual species. In this appendix, we will explain how do we compute particle densities based on the superficial densities of each species. In order to do so, we are going to suppose that the equation of state of the gas is:  
\begin{equation}\label{Eq:State} 
\rm	P_j = n_j \ K_B \ (T_j + T_{eff})
\end{equation}
where $\rm P_j$ is the partial pressure of the element j, $\rm n_j$ is its particle density, $ \rm K_B$ is Boltzmann constant, $\rm T_j$ is the temperature of the phase in which is immersed the species and $\rm T_{eff}=1000 K$ is an effective temperature that tries to reproduce the effects of turbulence and magnetic fields. As a consequence, we will estimate the total pressure of all the ISM using the expression that \citep{Pressure_of_a_disk_1,Pressure_of_a_disk_2} have obtained for a disk with axial symmetry: 
\begin{equation}\label{Eq:Pressure}
\rm	P \approx \frac{\pi}{2} \Sigma_{\rm gas}(\Sigma_{\rm gas} + \frac{\sigma_{\rm gas}}{\sigma_{*}} \Sigma_{*} + \Sigma_{d}) 
\end{equation} 
where $\frac{\sigma_{\rm gas}}{\sigma_{*}} \approx 1 $ is the ratio of velocity dispersion between the gas and the stars (Leroy et al. 2008). Therefore, the Eq. \eqref{Eq:Pressure} is transformed into :
\begin{equation}
	\rm P = \frac{\pi}{2} \ G \ \Sigma_{\rm gas} \Sigma_{total}
\end{equation}

However, we want the partial pressure of each species not the total pressure so knowing that $\rm P_j = X_j P$ where $\rm X_j \equiv \frac{\Sigma_j}{\Sigma_{\rm gas}}$  the partial pressure of the element j is:
\begin{equation}\label{Eq:partial_pressure}
\rm	P_j = \frac{\pi}{2} \ G \ \Sigma_j \Sigma_{total}
\end{equation}
Then, combining the Eq. \eqref{Eq:State} with the Eq. \eqref{Eq:partial_pressure}, the particle density is:
\begin{equation}
\rm	n_j = \frac{\pi \ G \Sigma_j  \Sigma_{total}}{2 \ K_B \ (T_j + T_{eff})}
\end{equation}

\bsp	
\label{lastpage}
\end{document}